\documentclass[journal]{vgtc}                     

\onlineid{1563}

\vgtccategory{Research}

\vgtcpapertype{Analytics \& Decisions}

\title{SKG: A Versatile Information Retrieval and Analysis Framework for Academic Papers with Semantic Knowledge Graphs}
\author{%
  Yamei Tu, Rui Qiu, Han-Wei Shen
}

\authorfooter{
  \item
  	Josiah Carberry is with Brown University.
  	E-mail: jcarberry@example.com
  \item
  	Ed Grimley is with Grimley Widgets, Inc.
  	E-mail: ed.grimley@example.com.

  \item Martha Stewart is with Martha Stewart Enterprises at Microsoft
  Research.
  	E-mail: martha.stewart@example.com.
}

\abstract{%
The number of published research papers has experienced exponential growth in recent years, which makes it crucial to develop new methods for efficient and versatile information extraction and knowledge discovery.  To address this need, we propose a Semantic Knowledge Graph (SKG) that integrates semantic concepts from abstracts and other meta-information to represent the corpus. The SKG can support various semantic queries in academic literature thanks to the high diversity and rich information content stored within. To extract knowledge from unstructured text, we develop a Knowledge Extraction Module that includes a semi-supervised pipeline for entity extraction and entity normalization. We also create an ontology to integrate the concepts with other meta information, enabling us to build the SKG. Furthermore, we design and develop a dataflow system that demonstrates how to conduct various semantic queries flexibly and interactively over the SKG. To demonstrate the effectiveness of our approach, we conduct  the research based on the visualization literature and provide real-world use cases to show the usefulness of the SKG. 
 The dataset and codes for this work are available at \url{https://osf.io/aqv8p/?view_only=2c26b36e3e3941ce999df47e4616207f}.

}

\keywords{Text analysis, Knowledge Graph, Natural Language Processing, Visual Interactive Query, Visual Analytics}

\teaser{
  \centering
  \includegraphics[width=\linewidth]{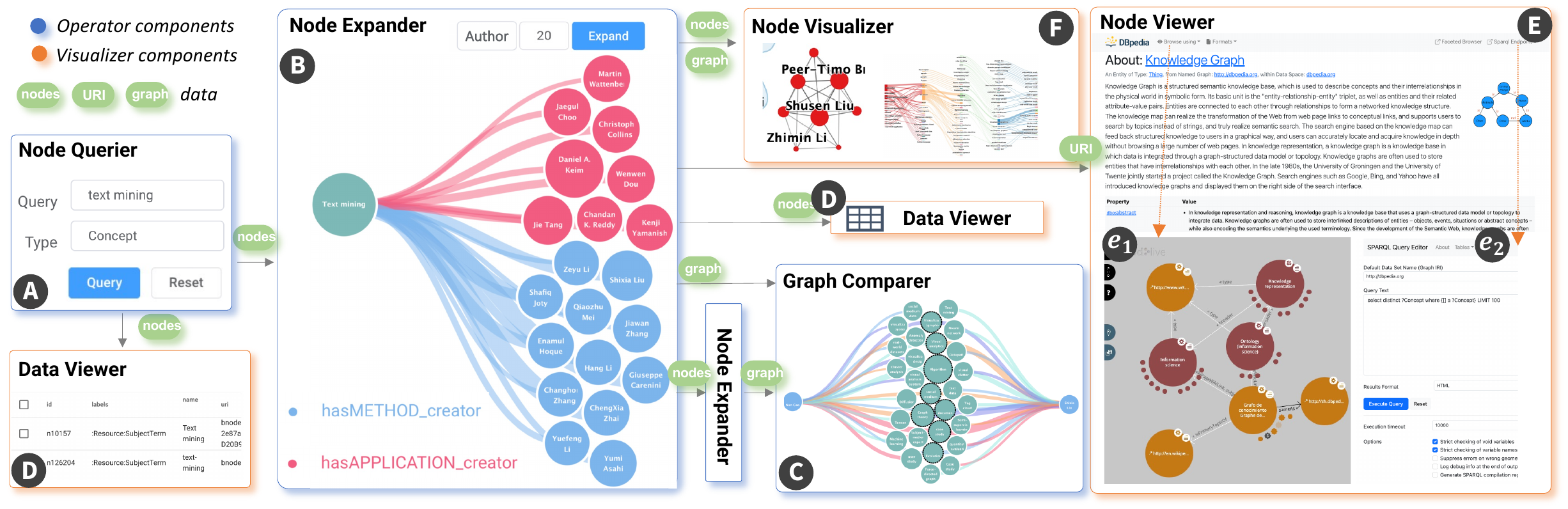}
\caption{Our Semantic Knowledge Graph is equipped with a dataflow system that allows users to create semantic queries in an interactive manner. The system includes operator components (\tikzcircle[fill={rgb,255:red,67; green,114; blue,196}]{4pt}) and visualizer components (\tikzcircle[fill={rgb,255:red,237; green,125; blue,49}]{4pt}), which users can connect using a simple  drag-and-drop interface to build various analysis pipelines for their specific requirements. }
\label{fig:system}
}

\graphicspath{{figs/}{figures/}{pictures/}{images/}{./}} 

\usepackage{tabu}                      
\usepackage{booktabs}                  
\usepackage{lipsum}                    
\usepackage{mwe}                       
\usepackage{amsfonts}
\usepackage{multicol}
\usepackage{multirow}
\usepackage{colortbl}
\usepackage{algpseudocode}
\usepackage{algorithm}
\usepackage{wrapfig}
\newcommand{\DATABASE}{\textit{VISBank}}

\newcommand{\blue}{\textcolor{black}}
   \newcommand{\centered}[1]{\begin{tabular}{l} #1 \end{tabular}}
\newcommand{\tikzcircle}[2][red,fill=red]{\tikz[baseline=-0.5ex]\fill[#1,radius=#2] (0,0) circle ;}%
\newcommand{\green}{\textcolor{teal}}

\newcommand*\circled[1]{\tikz[baseline=(char.base)]{
            \node[shape=circle,draw,inner sep=0.5pt] (char) {#1};}}

\makeatletter
\providecommand\underarrow@[3]{%
  \vtop{\ialign{##\crcr$\m@th\hfil#2#3\hfil$\crcr
  \noalign{\nointerlineskip\kern.12\baselineskip}#1#2\crcr}}}

  \providecommand{\underleftarrow}{%
  \mathpalette{\underarrow@\leftarrow@}}
\providecommand\leftarrowfill@{\arrowfill@\relbar\relbar\leftarrow}
\providecommand\arrowfill@[4]{%
$\m@th\thickmuskip0mu\medmuskip\thickmuskip\thinmuskip\thickmuskip
   \relax#4#1\mkern-7mu%
   \cleaders\hbox{$#4\mkern-2mu#2\mkern-2mu$}\hfill
   \mkern-7mu#3$%
}

\providecommand{\underrightarrow}{%
  \mathpalette{\underarrow@\rightarrowfill@}}
\providecommand\rightarrowfill@{\arrowfill@\relbar\relbar\rightarrow}
\providecommand\arrowfill@[4]{%
  $\m@th\thickmuskip0mu\medmuskip\thickmuskip\thinmuskip\thickmuskip
   \relax#4#1\mkern-7mu%
   \cleaders\hbox{$#4\mkern-2mu#2\mkern-2mu$}\hfill
   \mkern-7mu#3$%
}
\makeatother

\begin{document}
\maketitle
\section{Introduction}
Retrieving information from a large corpus is crucial, particularly in the academic context, where the goal is to facilitate the exploration and discovery of knowledge from scholarly literature~\cite{shah2002information,mayfield2003information}. Due to the recent explosion in the volume of publications, a single query can now generate a long list of documents, as seen in using search engines such as Google Scholar to find papers~\cite{ding2022tell}. It is also inefficient to manually read through all the discovered papers and summarize them to further research. If documents can be translated into a machine-processable format, it becomes possible to automate the search and summarization of documents by leveraging machine analytics power. 
The goal of this paper is, therefore, to develop solutions for automatic information retrieval from research documents to enable knowledge discovery. 

There are several challenges associated with existing technologies that must be addressed. First, transforming unstructured documents into structured formats is a complex task. While keywords are often used to represent the essential ideas of a document, a mere collection of keywords without proper organization cannot support semantic query. Topic modeling is another way to organize documents into clusters by identifying topics~\cite{jelodar2019latent}, but it is difficult to guarantee the quality of topics, and it requires human explanation~\cite{murdock2015visualization,sievert2014ldavis}. Such hidden topics are hard to utilize for accurate information retrieval.
Secondly, semantic queries require semantic understanding and knowledge distillation, which is not a trivial task. Although recent developments in Natural Language Processing (NLP) have led to the development of question-answering (QA) models to answer questions based on a contextual document~\cite{wang2019multi,allam2012question}, applying them to a large collection of documents is time-consuming and does not guarantee an organized output. Additionally, some sophisticated queries may involve both semantic and meta-data information, such as ``what are the research trend and the evolution path of OpenAI.'' This requires finding the publications from OpenAI and summarizing them, which can not be solved easily by current QA models.

 To tackle these challenges, we propose an approach that integrates Natural Language Processing (NLP) techniques with visual analytics (VA) to achieve interactive information retrieval. 
Our approach involves building a Semantic Knowledge Graph (SKG) as a fundamental representation of a large corpus and developing a dataflow system to support flexible semantic queries over SKG. 
The word ``Semantic'' has two levels of meaning in our approach. First, it refers to the concepts and their ``semantic'' relations from paper abstracts, which goes beyond many existing academic knowledge graphs that only incorporate meta-information.
For example, by pre-processing the unstructured text, our SKG contains ``topic modeling'' as a ``method'' while ``bioinformatic'' as an ``application''. 
Second, SKG contributes to the vision of the ``Semantic Web'' that creates a web of data that is machine-readable and interoperable across different applications and platforms. To support the Semantic Web, computers must have access to the structured information and sets of inference rules that they can use to conduct automated reasoning~\cite{berners2001semantic}. 
Combining the semantics with other meta-information results in the heterogeneity of our SKG, which can be utilized in two ways.
Firstly, by combining concepts with other meta-information, we can achieve flexible retrieval for different types of entities. Secondly, by summarizing connected semantic entities in a clear manner, our SKG can achieve automatic summarization of target entities, such as summarizing a list of papers for automatic literature review. It can be used to retrieve and summarize information efficiently. 

Building SKG can be a challenging task, particularly when it comes to extracting semantic entities from raw paper text. To address this, we propose a module that performs Named Entity Recognition (NER) and entity normalization.
NER is achieved by a semi-supervised approach that
leverages both machine and human efforts. It includes an unsupervised learning stage to generate weak training samples and a human fine-tuning stage to create a high-quality training dataset. The NER model is then fine-tuned on this dataset and can infer all named entities in the entire corpus. 
Second, to ensure that SKG contains accurate and non-redundant information, we also perform entity normalization by mapping the extracted entities to acknowledged knowledge using the DBpedia Spotlight API.

SKG can be used for semantic queries between different entity types. We have identified a set of tasks and usage scenarios that can be achieved by building queries from SKG. To make it easier for users to interact with SKG, we have divided these tasks into three key graph operations and visualization needs and embedded them into the components of a dataflow system. Each component has been designed with a single function with specific inputs/outputs, enabling users to drag and drop these components to build different analysis pipelines, providing flexibility to fulfill various tasks.
We demonstrate the usefulness and effectiveness of our approach for exploring the visualization literature by collecting 125,745 papers, which we refer to as the $\DATABASE$ dataset. We then construct SKG over the $\DATABASE$ dataset and use our dataflow system to query SKG built from the visualization literature.
 In summary, our contributions are:
\begin{itemize}[leftmargin=*]\setlength\itemsep{-0.5em}
    \item The design and construction of a semantic knowledge graph to enable efficient semantic queries for large academic corpora.
    \item The construction of a semi-supervised knowledge extraction module for named entity recognition and entity normalization.
    \item The design and development of a dataflow system that enables users to perform semantic queries over SKG using visual operations.
    \item The publication of the $\DATABASE$ and SKG as a contribution to further research in the visualization literature.
\end{itemize}

\section{Related Work}

Knowledge Graphs (KG) have attracted significant attention from both academia and industry due to their superior performance in data integration. To benefit the public community, several public KGs have been proposed, including DBpedia~\cite{auer2007dbpedia}, Freebase~\cite{bollacker2008freebase}, YAGO~\cite{suchanek2007yago}, and MAKG~\cite{wang2020microsoft}, to name a few. 
We discuss KGs from two perspectives that are most related to our work: the construction of KGs and the visualization system for KGs. 
\subsection{Construction of Knowledge Graph}
In the current era of big data, knowledge modeling is an effective way to collect and synthesize knowledge scattered in mass data. 
It can be divided into two main techniques: ontology-based and non-ontology approaches. 
Non-ontology approaches extract entities and relationships from unstructured text without relying on pre-defined ontology. 
Yun et al.\cite{yun2021knowledge} provide a comprehensive summary of applicable methods, involved processes, and techniques for the non-ontology approach. 
Some existing works reply on state-of-the-art language processing tools~\cite{buscaldi2019mining, dessi2020ai, dessi2021generating}, such as Extractor Framework~\cite{luan2018multi}, CSO Classifier~\cite{salatino2019cso}, OpenIE~\cite{angeli2015leveraging}, CoreNLP~\footnote{\url{https://stanfordnlp.github.io/CoreNLP/pos.html}} to extract knowledge. Others propose their own models to extract relations and entities for KG construction.
Chansanam et al~\cite{chansanam2022culture} use a BiLSTM model to construct a Thai cultural knowledge graph.
Guo et al.~\cite{guo2022automatic} apply the function-behavior-states (FBS) design method to classify and extract knowledge using the BERT-BiLSTM-CRF model, and 
 Feng et al.~\cite{feng2021small} build an automatic information extraction model based on the BiLSTM-CRF model to extract knowledge from accident reports, laws, and regulations. 

Ontology-based methods are the most common and well-studied, which is also the focus of our work. 
Mondal et al.~\cite{mondal2021end} define an ontology of NLP-KG and propose an end-to-end framework that contains three different relation extractors to identify pre-defined relation types. 
Al-Khatib et al.\cite{al2020end} define an argumentation KG and propose a supervised approach to detect relations, while syntactic patterns identify entities. 
We also found similar efforts to construct KGs in scientific literature.
Chen et al.\cite{chen2019automatic} proposed an ontology-based pipeline to build KGs from abstracts. However, they perform sentence classification first and then extract entities from each sentence, which cannot be applied in scenarios where  entities with different labels can co-occur in the same sentence. 
Tosi et al.~\cite{tosi2021scikgraph} aim to structure knowledge in scientific literature, but they use a Babelfy~\cite{moro2014entity} to extract entities and map them to BabelNet~\cite{navigli2012babelnet}, which is a knowledge graph built on WordNet~\cite{miller1995wordnet}. 
Wise et al.~\cite{wise2020covid} construct  an ontology for COVID-19 from CORD19 and extract entities using SciSpacy, which has been trained on biomedical texts similar to CORD19. 
However, due to the large volume of scientific publications,  a dictionary database or domain-specific extractor cannot be guaranteed to exist for all scientific applications, such as visualization. Therefore, in this work, we propose a framework that does not rely on pre-existed materials.





\subsection{Visualization of Knowledge Graph}
The development of KGs also highlights the need for effective data visualization. Nararatwong et al.~\cite{9355442} discussed several challenges of visualization KGs due to their massive and complex nature, while Gomez et al.~\cite{gomez2018visualizing} conducted a performance analysis on various visualization tasks for large-scale knowledge graphs. 

There are many existing efforts on building visualization systems for KGs, with various focuses. 
Due to the steep learning curve of structured query languages, several visualization systems enable users to query KGs using visual query formulations (e.g., OptiqueVQS~\cite{soylu2018optiquevqs}) or graph-like queries (e.g., RDF Explorer~\cite{vargas2019rdf}, FedViz~\cite{e2015fedviz}).

As querying is the first step during KG exploration, some tools support visualizations that efficiently represent queried data. Depending on the targeted KGs, these systems can be categorized as general or domain-specific solutions. 
Deagen et al.~\cite{deagen2022fair} defined a chart as a combination of SPARQL query and Vega-Lite grammar for chart specification to represent information in Scientific KGs while Wei et al.~\cite{wei2020vision} proposed a topic-centric visualization system to summarize and cluster neighboring entities for query entities. CEPV ~\cite{sheng2019cepv} is a customized information extraction, process and visualization tool to extract data from KG into a specified tree structure visualization. 
Several domain-specific visualization systems have been built for various purposes~\cite{pellegrino2020move, hirsch2009interactive, 8731539}. Qiu et al.~\cite{qiu2020tax} developed a taxation big data visualization system for tax-related problems while Jiang et al.~\cite{9239940} proposed a visual analysis system to perform spatiotemporal analysis of a COVID-19 knowledge graph. 
ALOHA~\cite{he2019aloha},  a dietary supplement knowledge graph visualization developed for exploring iDISK knowledge base. 
To the best of our knowledge, we propose the first dataflow system for KGs, which enables flexible  information retrieval and analysis  in scientific literature. The system design is easily adaptable for different KGs, making it data-agnostic. 
\section{Method}
Our approach to enhancing information retrieval and analysis in the academic literature combines natural language processing (NLP) techniques with visual analytics. 
In this section, we first introduce our Semantic Knowledge Graph (SKG) and then discuss: (1) how we construct SKG in a semi-supervised manner; (2) how SKG empowers the semantic queries in academic literature; (3) how to enable users to construct queries interactively and discover insights efficiently. 
\subsection{Semantic Knowledge Graph}\label{sec:skg}

The Semantic Knowledge Graph (SKG) serves as a foundational representation of a large corpus.
It is a graph-based representation of knowledge that captures entities and their relationships. 
Unlike some existing academic knowledge graphs, SKG combines information from both meta-data and text data. 
As shown in \autoref{fig:ontology}, SKG has five types of entities, one \textbf{semantic entity}, called \emph{Concept}, and the other four types of \textbf{meta-data entity}.
Each entity type has a specific meaning and related attributes, summarized as follows:
\begin{figure}[htp]
    \centering
    \includegraphics[width=7cm]{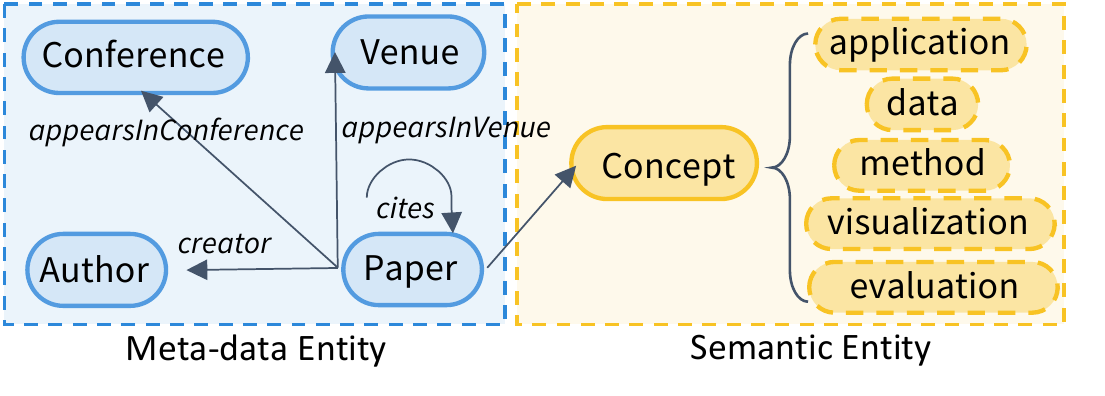}
    \vspace{-3mm}
    \caption{The schematic ontology of Semantic Knowledge Graph. It contains four meta-data entity types and one semantic entity type. \emph{Concepts} with different semantic roles are captured in relationships between \emph{Paper} and \emph{Concept}.}
    \label{fig:ontology}\vspace{-3mm}
\end{figure}
\begin{itemize}[leftmargin=*]
    \item \emph{Paper}: Represents each paper with attributes of title, a unique id, publication year, and URL to the semantic scholar. It has outgoing relations to all the other four entity types. 
  \item \emph{Concept}: Represents the scientific keyword or keyphrases extracted from paper abstracts.
  If the \emph{Concept} is recognized in the DBPedia, it will include a URL attribute that leads to the corresponding DBPedia page. 
  Since \emph{Concept} can have different semantic roles in abstracts, we capture such information in the relationships that connect \emph{Concept} and \emph{Paper}. Specifically, we define five types of \emph{Concept} to categorize knowledge in abstracts:
  \begin{itemize}[leftmargin=*]
      \item \emph{Application}: the \emph{Concept} that describes the \emph{application} such as biology, business, or  \emph{task} like image classification.
      \item \emph{Data}: the \emph{Concept} that describes the \emph{data} or \emph{dataset}, e.g., volume data, image,  MNIST dataset.
      \item \emph{Method}: the \emph{Concept} that describes the \emph{method}, \emph{algorithm} or \emph{technique}, such as cross validation, Random Forest.
      \item \emph{Visualization}: the \emph{Concept} that describes the \emph{visualization} or \emph{chart}, e.g., visual analytic system, bar chart, heat map.
      \item \emph{Evaluation}: the \emph{Concept} that describes the \emph{evaluation}, e.g., quantitative evaluation, expert interview, user study.
      
  \end{itemize}
    \item \emph{Author}: Represents authors with an attribute of the full name. 
    \item \emph{Journal}: Represents journal entity with an attribute of the name. 
    \item \emph{Venue}: Similar to \emph{Journal} entity, it represents the venue where one paper has been published. 
\end{itemize}


\subsection{SKG Construction}
 \blue{Constructing a useful and effective knowledge graph can be challenging, as it depends on both the quantity and quality of data it contains. Therefore, as the first step, we create a large domain-specific dataset that includes both metadata and raw text to be used in the SKG. We then extract concepts from the raw text, to transform the unstructured textual data into structured knowledge.
 To accomplish this, we develop a semi-supervised module that combines machine computation ability with human intelligence. 
Finally, using the metadata and concepts obtained from the previous steps, we represent our knowledge graph in the Resource Description Framework (RDF). The RDF format is the standard for representing knowledge in the semantic web, which allows machines to process and reason about data in a meaningful way.
In summary, our SKG construction involves the following steps:
\begin{itemize}[leftmargin=*]\setlength\itemsep{-0.5em}
    \item VISBank Dataset: We create a large and rich academic dataset in the visualization literature for knowledge graph construction. 
    \item Concept Entity Extraction: We build a semi-supervised framework to extract concepts and their roles from unstructured text,  to be incorporated in the SKG. 
    It also contains a normalization stage to remove duplicate knowledge. 
    \item RDF representation: We describe the statistics of constructed SKG and provide concrete examples. 
\end{itemize}}
\noindent Below, we describe each of the steps in detail. 
\vspace{-1mm}

 \subsubsection{$\DATABASE$ dataset}
\blue{Several well-known academic datasets are available in the visualization domain. However, they do not cover all the necessary information we need as they are not built for knowledge graph purposes. 
For example, VisPub~\cite{isenberg2016visualization} is a popular benchmark dataset that contains various metadata about papers in the IEEE VIS conferences, 
but it does not cover other relevant fields, such as virtual reality, or graphics. 
Another dataset, VitaLITy~\cite{narechania2021vitality}, covers 38 popular data visualization publication venues. However, it does not contain the citation relationships as it focuses on utilizing abstract to 
generate semantic embeddings for serendipity discovery. 
Thus, we construct a new dataset called $\DATABASE$, which is ideal for constructing knowledge graphs in the visualization domain.  }

\begin{table}[htp]
\small
\caption{Statistical information regarding the $\DATABASE$ dataset includes (A) the coverage of key attributes of papers covered in $\DATABASE$; (B) a comparison of the $\DATABASE$ with other datasets in the VIS literature; (C) The paper distributions across sub-areas in the VIS literature. }
\setlength{\tabcolsep}{1pt}
\begin{tabular}{l|lr|l|lr}
\hline
\multirow{5}{*}{\textbf{\begin{tabular}[c]{@{}l@{}}(A)\\ Coverage\\ of Key\\ Attribute\end{tabular}}} & Title & 100\% & \multirow{8}{*}{\textbf{\begin{tabular}[c]{@{}l@{}}(C)\\ Sub\\ Area\end{tabular}}} & Graphics \&Visualization & 55,759 \\
 & Abstract & 85.2\% &  & Human-Computer Interaction & 36,719 \\
 & Author & 99.7\% &  & Knowledge \& Data mining & 9,490 \\
 & Venue/Journal & 99.3\% &  & Virtual/Augmented Reality & 7,648 \\
 & Total & 125,745 &  & Cognitive Visualization & 6,514 \\ \cline{1-3}
\multirow{3}{*}{\textbf{\begin{tabular}[c]{@{}l@{}}(B)\\ Compare\end{tabular}}} & VisPub & 3,120/3,503 &  & Mobile \& Ubiquitous Visualization& 4,524 \\
 & VitaLITy & 52,784/59,232 &  & User Interface, System & 2,279 \\
 & S2ORC & 125,745/135M &  & Others & 2,812 \\ \hline
\end{tabular}
\label{tab:visbank}
\vspace{-3mm}
\end{table}

\blue{$\DATABASE$ offers high extensiveness and diversity in terms of the data it covers, making it a valuable resource for researchers  in the visualization field.
The dataset contains $125,745$ papers from $3520$ venues and $88$ journals, following the same format as the Semantic Scholar Open Research Corpus (S2ORC)~\cite{lo-wang-2020-s2orc}~\footnote{\url{https://github.com/allenai/s2orc}}. 
We compute the coverage of several key attributes in the $\DATABASE$ and report it in \autoref{tab:visbank}A. The high percentage indicates that $\DATABASE$ provides comprehensive metadata coverage, providing a strong foundation for SKG construction. 
Additionally, we compare $\DATABASE$ with two aforementioned datasets, i.e., \emph{VisPub} and \emph{VitaLITy}. The results are presented in \autoref{tab:visbank}B. The high coverage of both datasets suggests that our $\DATABASE$ merges various data sources in the visualization literature, ensuring compliance with the FAIR principles for making data more \textbf{f}indable, \textbf{a}ccessible, \textbf{i}nteroperable, and \textbf{r}eusable. 
Also, the paper distribution across related areas in the visualization literature is shown in \autoref{tab:visbank}C, highlighting the multidisciplinary nature of the data in the $\DATABASE$.}
\begin{figure*}[htp]
    \centering
    \vspace{-2mm}
    \includegraphics[width=\linewidth]{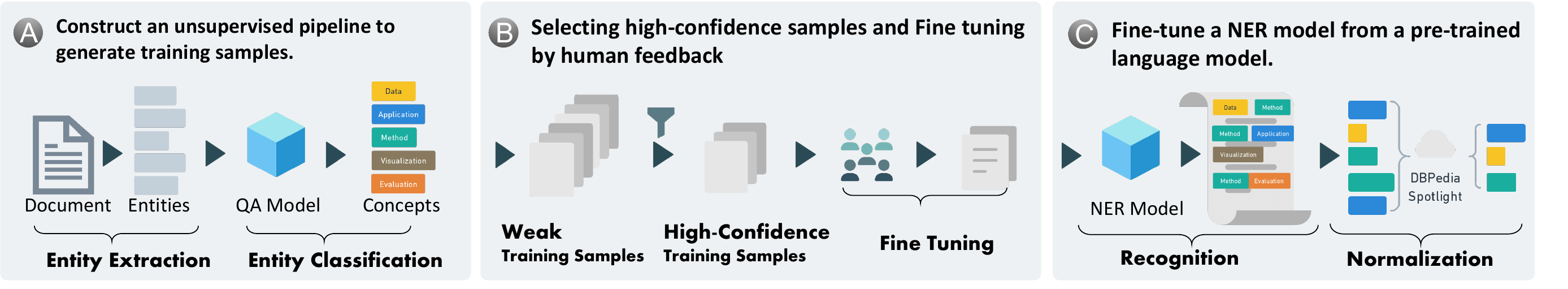}
    \caption{The process of transforming unstructured text into \emph{Concept} entities involves three steps: (A) preparing a training dataset for the Named Entity Recognition (NER) model by utilizing machine semantic understanding ability, (B) incorporating human feedback into the training dataset, and (C) training the NER model to recognize all entities in the $\DATABASE$ and perform normalization to eliminate redundant information. }
    \label{fig:ner}
    \vspace{-6mm}
\end{figure*}
\blue{We construct the $\DATABASE$ dataset by extracting relevant papers from the S2ORC dataset. The reason is that S2ORC contains over 136 million scientific papers with rich meta-data, including raw text information. 
However, it can be challenging to identify a small  target subset from such a large corpus. To narrow down the selection, we first filtered the attribute of ``field of study'' to ``computer science'', which is the smallest superset of visualization, resulting in 12 million papers.  Next, we identify the relevant sub-fields by performing a keyword-based search. For instance, we use the keyword ``visualization" to identify related journal and venue names such as \emph{Journal of Visualization} and \emph{IEEE Transactions on Visualization and Computer Graphics}. By using this approach, we have compiled a full list of $3520$ relevant venues and $88$ journals, which can be found in the supplemental material. We further filter papers that appear in those venues and journals. 
Additionally, we also map papers from \emph{VisPub} and \emph{VitaLITy} to S2ORC by matching their titles and integrate the matched ones into $\DATABASE$. We process each abstract to extract \emph{Concept} entities as explained below and add this information into $\DATABASE$. }

\subsubsection{Concept Entity Extraction}

$\DATABASE$ contains meta information that can be directly transformed into the meta-data entities in SKG. 
However, the \emph{Concept} entity needs to be extracted and classified from the abstracts, which can be a challenging task when considering the requirement of high accuracy and efficiency.
To address this issue, we propose a semi-supervised module to perform the Named Entity Recognition (NER) task that simultaneously recognizes entities and classifies them into the five named \emph{Concept} types mentioned above. 
The module consists of three key steps to reduce the labeling effort required to prepare the training dataset for NER: (1) unsupervised pipeline to generate weak training samples; (2) filtering high-confidence samples and soliciting human annotators to create high-quality training samples; (3) supervised training of a NER model, as shown in \autoref{fig:ner}.

To generate weak training samples, our unsupervised pipeline first extracts candidate entities from the text. 
To do so, we first generate the Part-of-Speech (POS) tag for each token, indicating its grammatical function in the sentence. We then use a rule-based approach to extract the noun n-grams based on the POS tags. This approach has high readability and is easy to tune to achieve high coverage of all candidate entities~\cite{waltl2018rule}. We adopt the same rules from previous work~\cite{tu2022phrasemap}.

Next, we classify the candidate entities into the five categories of \emph{Concept}. 
Recent findings in the NLP literature suggest that language models have a certain level of semantic understanding~\cite {maulud2021state,shen2023chatgpt}. Inspired by this, we utilize a Question-Answering (QA) model to generate labels for entities based on their contextual information. 
 The high-level idea is to translate each \emph{Concept} label into questions using the definition provided in \autoref{sec:skg}. 
 For example, the \emph{data} label is translated into two questions: \textit{what is the data?} and \textit{what is the dataset?}
 Then, we assign each entity to one label with the highest possibility of answering its corresponding question. 
For example, 
``volume data'' has the highest probability of answering the question of \textit{what is the data?} and therefore will be assigned to the \emph{data} label.


\blue{For question answering, we employ a RoBERTa model
that is fine-tuned on the SQuAD2.0~\cite{2016arXiv160605250R} QA dataset.
Traditionally, a QA model takes a question and a document $(q, d)$ as input and extracts a span of tokens from the document as the answer. 
The model uses two sets of trainable weights, $\textbf{W}_S, \textbf{W}_E$, to compute the score of each token $t_i$ to be the start or end position of the answer as follows:}
\begin{equation}\small
      score_S(q, d, t_i) = \frac{e^{{\textbf{W}_S} \cdot \bf{v}_i}}{\sum_{j}e^{{\textbf{W}_S} \cdot  \bf{v}_j}}, \quad score_E(q, d, t_i) = \frac{e^{{\textbf{W}_E} \cdot \bf{v}_i}}{\sum_{j}e^{{\textbf{W}_E} \cdot \bf{v}_j}}.
\end{equation}\blue{ $\textbf{v}_{i}$ is the embedding of a token $t_i$, which is an intermediate output in the model. $\textbf{v}_i$ shares the same dimensions with the weight matrix $\textbf{W}_S,\textbf{W}_E$ ($768d$). Then, the probability of a span $s$ starting from position $i$ and ending at position $j$ to answer the question is computed as: 
\begin{equation}\label{eq:qa}
    p(s_{t_i:t_j}|q, d) = Softmax(score_S(q,d,t_i)\cdot score_E(q,d,t_j))
\end{equation} where $p(s_{t_i:t_j}|q, d)$ is a probability distribution of all possible spans. Then the span with the highest probability is returned as the answer. 
}

\blue{
To apply the QA model to our use case, for each candidate entity $c$, we compute its probability of answering each question $q_m$ using \autoref{eq:qa}.
Then we find the question with the highest probability and 
assign its corresponding to the candidate entity $c$.
After processing all documents $d$ in $\DATABASE$, we obtain a set of (document $d$, candidate entity $c$, label $l$, probability $p$) as \textbf{weak} training samples, where each tuple indicates one candidate entity $c$ in $d$ being assigned to label $l$ with the probability of $p$. For each label $l$, we sort the corresponding tuples based on the probability $p$ and select top-K as \textbf{high-confidence} training samples, as shown in \autoref{fig:ner}B}.
 
 
 \begin{table*}[]
\small
\centering 
\caption{Summarizing the utilization of our SKG in different real-world scenarios that involve semantic queries. }
\begin{tabular}{llll}
\hline
\multicolumn{1}{c}{Entity} & \multicolumn{1}{c}{Ontology} & \multicolumn{1}{c}{Supported Task}  & \multicolumn{1}{c}{Usage Scenarios of Semantic Queries} \\ \hline
Paper & \centered{\includegraphics[width=1.6in]{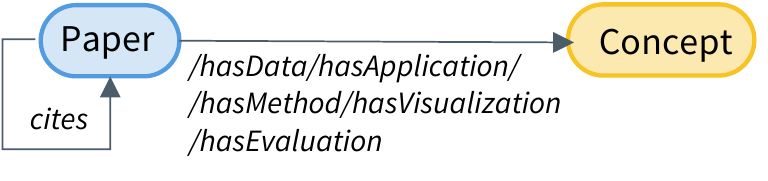} }& 
\begin{tabular}[c]{@{}l@{}}• Document Retrieval\\ • Automatic Literature \\Review\end{tabular} & \begin{tabular}[c]{@{}l@{}} Find academic papers related to \emph{\underline{text mining}} and \emph{\underline{visualization}}, and provide a  summary of the \\\emph{\underline{tasks}} that can be addressed by the different \emph{\underline{methods}}  used in these papers.  \end{tabular}  \\ \hline
Author & \centered{\includegraphics[width=1.6in]{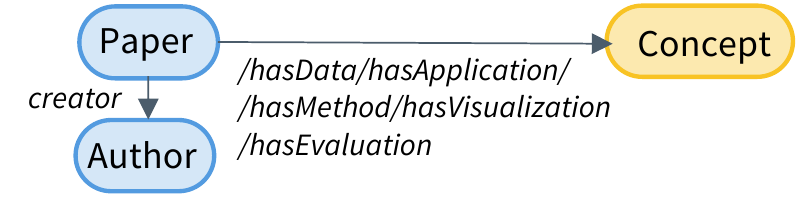}} &  
\begin{tabular}[c]{@{}l@{}}• Reviewer Finder\\ • Researcher Profiling \end{tabular} 
& 
\begin{tabular}[c]{@{}l@{}}
Find potential reviewers to evaluate a new research paper focused on using \underline{\emph{neural networks}}\\ for \emph{\underline{parameter exploration in ensemble}}  \underline{\emph{simulations}}.
\end{tabular}
\\  \hline
\begin{tabular}[c]{@{}l@{}}Venue\\ /Journal\end{tabular} & 
\centered{\includegraphics[width=1.7in]{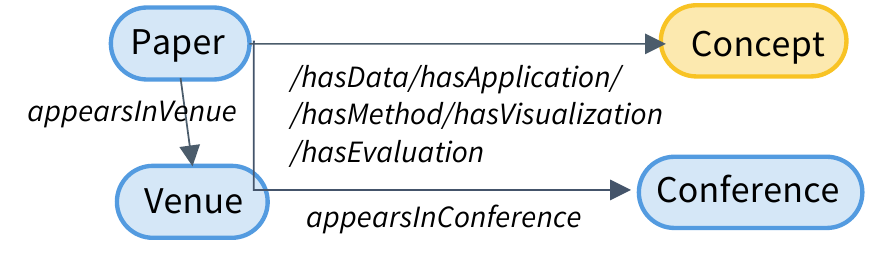}}
&\begin{tabular}[c]{@{}l@{}}• Conference Finder\\ • Journal Comparison\end{tabular}  
& \begin{tabular}[c]{@{}l@{}}
Find possible conferences to submit the work related to \emph{\underline{improving trust in ML }} \\\emph{\underline{through visualization}} and compare those conferences based on their accepted papers. 
\end{tabular}
\vspace{-1mm}
\\ \hline
\end{tabular}
\label{tab:query}
\vspace{-4mm}
\end{table*}

We manually refine weak training samples as they may not be perfect.
The annotation process is conducted iteratively. Initially, we labeled entities based on the \emph{Concept} definitions provided in \autoref{sec:skg}. 
For instance, we labeled entities related to tasks and applications as \emph{application}. However, during the human annotation phase, disagreements arose between the machine and the human annotators. To resolve these discrepancies, we held discussions with five PhD students who have expertise in visualization. We discussed and agreed upon a proper solution and applied it consistently throughout the annotation pipelines. 

We labeled a total of $5076$ samples and converted entity-level labels into a token-level format
to train the NER model. In this format, each token in a sequence $[t_1, t_2,.., t_i]$ is assigned a label $[l_1, l_2,.., l_i]$, where $l_i\in$\textit{\{application, data, method, visualization, evaluation\}}. 
In addition, the label $l_i$ for each token is prefixed with $B$ or $I$ to indicate whether the token is the beginning or continuation of a concept. $O$ indicates the token is not contained in any concept. 
For instance, given the sentence
``we conduct network analysis to graph data'', the corresponding labels for each token are $<$O, O, B\_method, I\_method, O, B\_data, I\_data$>$. 

The NER model is fine-tuned by computing the cross entropy between the ground truth token labels and predicted token labels using the cross-entropy loss, computed as follows:
\begin{equation} 
    H(P,Q) = -\sum_{t_i}{P(t_i)\cdot log(Q(t_i))}
    \vspace{-3mm}
\end{equation} where $P(t_i)$ is the actual distribution of token $t_i$ among all possible labels, while $Q(t_i)$ represents the predicted distribution.
The fine-tuned NER model can take a document as input and automatically recognize named \emph{Concept}.
We apply it to perform inference on the entire dataset and incorporate the results into the $\DATABASE$ dataset, which are used as \emph{Concept} entities in SKG.



\subsubsection{Semantic Knowledge Graph Construction}
Using the ontology defined in \autoref{fig:ontology}, we transform the metadata and semantic information from $\DATABASE$ into entities and build relationships between them to construct SKG. 
Additionally, we check whether each \emph{Concept} has a similar entity defined in DBpedia by using DBpedia Spotlight, as shown in \autoref{fig:ner}C. If a similar \emph{Concept} is found, we add the DBpedia URL as an additional attribute of the \emph{Concept} and  rename it using the recognized name in DBPedia. 
This process has two benefits: (1) it helps to normalize \emph{Concept} entities that have different wordings or descriptions but are actually referring to the same thing, and
(2) DBpedia provides detailed explanations for concepts as it links to Wikipedia. By integrating this information into SKG, we can enrich the knowledge embedded and provide users with \emph{Concept} explanations when they are exploring SKG.


SKG is represented using RDF (Resource Description Framework) triplets because it is the most powerful and expressive standard designed by the World Wide Web Consortium (W3C). RDF triplets harmonize all the relations into a $<s, p, o>$ format, where the subject entity $s$ and object entity $o$ are connected by the predicate $p$. 
This generality can integrate various types of information, which is efficient for data publication and interchange. 
The triplets follow the schematic ontology defined in \autoref{fig:ontology}, and a specific version of the schema can be found in the supplemental material. 

\begin{wrapfigure}{r}{0.25\textwidth}
\vspace{-6mm}
  \begin{center}
    \includegraphics[width=0.25\textwidth]{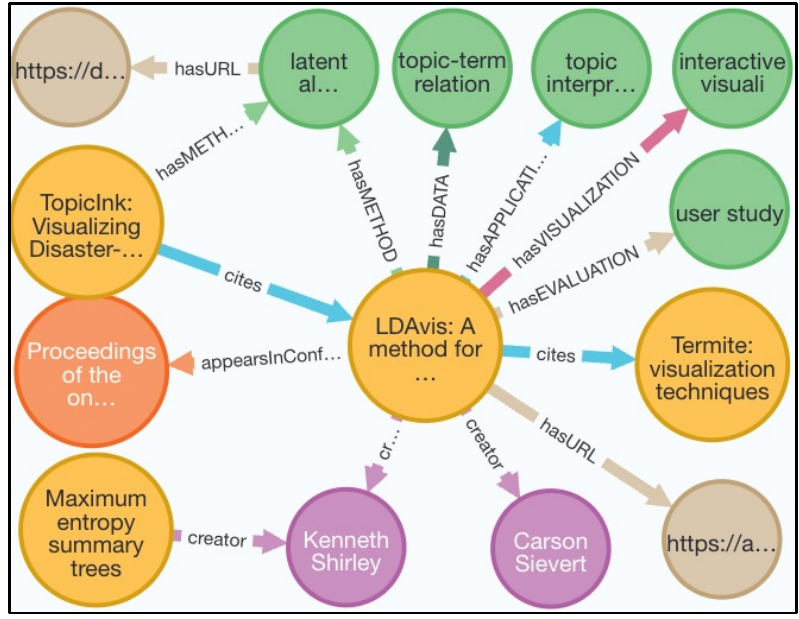}
  \end{center}
  \vspace{-5mm}
  \caption{A subgraph from SKG that is built from one paper record.}
  \vspace{-3mm}
   \label{fig:skg_example}
\end{wrapfigure}
After construction, each instance in $\DATABASE$ can be seen as a subgraph that surrounds the \emph{Paper} entity. 
An example of such a subgraph is shown in \autoref{fig:skg_example}. 
The paper titled ``LDAvis: a method for visualizing for visualizing and interpreting,'' was written by Carson Sievert, and Kenneth Shirley and published in \emph{Proceedings of the Workshop on Interactive Language Learning, Visualization, and Interfaces}.  
The paper focuses on the task of \emph{topic interpretation} based on \emph{topic-term relationship} using \emph{Latent Dirichlet Allocation} and \emph{interactive visualization}. 

\begin{table}[hbt]
\centering
\small
\caption{The statistics of different entities and relations in SKG. 
\emph{(Conf* is short for Conference)}}
\begin{tabular}{ll|ll}
\hline
\multicolumn{1}{c}{Entity Type} &
  \#Number &
  \multicolumn{1}{c}{Relation Type} &
  \#Number \\ \hline
\begin{tabular}[c]{@{}l@{}}Paper\\  \\ Concept\\ \\ Author\\ \\ Journal \\ \\ Conference\end{tabular} &
  \multicolumn{1}{r|}{\begin{tabular}[c]{@{}r@{}}125,745\\ \\ 766,276\\ \\ 181,031\\ \\ 1,355\\ \\ 4,367\end{tabular}} &
  \begin{tabular}[c]{@{}l@{}}cites (Paper$\rightarrow$Paper)\\ creator (Paper $\rightarrow$ Author)\\ appearsInJournal (Paper$\rightarrow$Journal)\\ appearsInConf* (Paper$\rightarrow$Conf*)\\ hasData (Paper$\rightarrow$Concept)\\ hasApplication (Paper$\rightarrow$Concept)\\ hasMethod (Paper$\rightarrow$Concept)\\ hasVisualization (Paper$\rightarrow$Concept)\\ hasEvaluation (Paper$\rightarrow$Concept)\end{tabular} &
  \multicolumn{1}{r}{\begin{tabular}[c]{@{}r@{}}344,389\\ 398,670\\ 57,330\\ 111,558\\ 350,262\\ 333,562\\ 537,921\\ 99,284\\ 172,981\end{tabular}} \\ \hline
\end{tabular}
\label{tab:skg}
\end{table}


\blue{\autoref{tab:skg} presents the counts of various entity types and relation types contained in the constructed SKG. Each of the $\sim$126k \emph{Paper} entities has an average of 3.1 outgoing relations to \emph{Author} entities and 2.7 outgoing relations that cite other papers. Additionally, our module identifies an average of 11.9 \emph{Concept} entities for each \emph{Paper}. Among these \emph{Concept} entities, around 23.4\% are related to data or datasets, 22.3\% pertain to applications or tasks, and 36\% refer to methods, algorithms, or techniques discussed in the abstract. Approximately 6.6\% entities are related to visualization since some papers do not explicitly discuss visualization techniques. Finally, 11.6\% entities introduce the evaluation methods or metrics employed.}

\subsection{Semantic Queries over the SKG}
Semantic queries over SKG can be seen as extracting a subgraph by following some search criteria.
In this section, we explain how SKG empowers information retrieval and analysis through semantic queries.
However, due to the complexity and diversity of the SKG, constructing different semantic queries to achieve various goals in structured query languages such as SPARQL can be a time-consuming task.
To overcome this challenge, we extract the essence of semantic queries and develop an algorithm that handles  various semantic queries with high efficiency and scalability.


\blue{
Some academic knowledge graphs stay at the metadata level, which can support a certain number of tasks that only need meta-information, such as citation analysis, and affiliation research tracking. However, advanced tasks usually require human intervention to process text, such as reading, understanding, and summarizing. 
This is why we put significant efforts into constructing the SKG, by extracting, summarizing, and integrating information from both textual content and metadata.
By incorporating both sources of information into a single knowledge graph, we can automate many real-world tasks that previously required human efforts. Later, we discuss advanced semantic queries that leverage both semantic  and meta-data entities. 
}

SKG is capable of supporting a wide range of queries by establishing rich connections among different entities through direct or multi-hop relations.
We categorize these queries into two different tasks: retrieval and summarization. 
In terms of retrieval, connections from semantic entities to other meta-data entities can enhance information retrieval efficiency. 
For example, to find potential reviewers of a submission discussing ``graph drawing'', starting from ``graph drawing'', we can 
retrieve authors who have published papers on ``graph drawing''  by traversing SKG.
This approach can provide factual data to support decision-making instead of relying on personal experience or domain knowledge.
On the other hand, the connections from meta-data entities to semantic entities are highly useful for information summarization. For instance, when dealing with multiple papers, automatic summarization from multiple perspectives is helpful, which can be achieved by utilizing various relationships between \emph{Paper} and \emph{Concept} entities.

Semantic queries share a common characteristic of identifying a set of starting entities and specifying search criteria to navigate SKG and reach the target entities. 
In \autoref{tab:query}, we summarize multiple queries based on the type of target entity.
The \emph{ontology} column displays the relevant entity and relation types utilized to support the query. The \emph{supported task} column contains two example tasks related to retrieval and summarization. We also provide  \emph{examples} for illustration purposes. 

\begin{algorithm}
\small
\caption{Semantic Query over the SKG}\label{alg:semanticquery}
\begin{algorithmic}[1]
\State \textbf{Input}: SKG $\mathcal{G}=(\mathcal{N}, \mathcal{E})$ and its ontology $\mathcal{G}_{ont}$, source entities $\mathcal{N}_s$,  number of target entities $K$, desired entity type $t$
\State \textbf{Output}: target entities $\mathcal{N}_t$ with type $t$, $\mathcal{E}_{s\rightarrow t}=\{n_i \rightarrow n_j|n_i\in\mathcal{N}_s, n_j\in\mathcal{N}_t\}$
\State \textbf{Declare}: Candidates nodes  $\mathcal{N}_c$, Candidate edges $\mathcal{E}_c$
\For{$n_i \in \mathcal{N}_s$}
\State $d$ $\gets$ \texttt{shortest\_dist}(\texttt{type}($n_i$), $t$, $\mathcal{G}_{ont}$) \green{// hops between source and target entity}
\State  $\mathcal{N}_d \gets$ \texttt{all\_destinations}($n_i$, cutoff=$d$, $\mathcal{G}$)  \green{// all reachable nodes with d-hops }
\For{$n_d \in \mathcal{N}_d$}
    \If{\texttt{type}($n_d$)= $t$} \green{// find candidate entities with type t }
    \State $s\gets \texttt{score}(n_i, n_d, \mathcal{G})$ \green{// the number of connections between ($n_i, n_d$) }
    \State $\mathcal{N}_c\gets \texttt{add}((n_d, s))$ \green{// keep recording score of each candidate entity}
    \State $\mathcal{E}_c\gets \texttt{add}((n_i, n_d))$ \green{// adding the relation}
    \EndIf
\EndFor
\EndFor
\State $\mathcal{N}_t\gets\texttt{sortedByValue}(\mathcal{N}_c)$[0:K] \green{// sort the candidate entities and return the top-K}
\For{$e=(n_i, n_j)\in \mathcal{E}_c$}
\If{$ n_j\in \mathcal{N}_t $} \green{// filter edges according to target entity}
\State $\mathcal{E}_{s\rightarrow t}\gets\texttt{add}(n_i, n_j)$
\EndIf
\EndFor
\end{algorithmic}
\end{algorithm}
Based on the commonality of queries,  we have designed a versatile algorithm to support various queries, as outlined in \autoref{alg:semanticquery}. 
The algorithm takes in source entities $\mathcal{N}_s$ as starting points in SKG, a desired target entity type $t$ as the search conditions, and $K$ as the number of returned entities. 
It will extract a subgraph from SKG, including source entities $\mathcal{N}_s$, $K$ most relevant target entities $\mathcal{N}_t$, and connections between the source and target entities ($\mathcal{E}_{s\rightarrow t}$).

Specifically, starting from each source entity $n_i$, we first find all reachable entities $n_d$ with type $t$ (\emph{line5-8}). 
Since $n_i$ may not be directly connected to $n_d$, such as \emph{Author} and \emph{Concept}, we count the number of all simple paths between $n_i$ and $n_d$ as a score for $n_d$ (line9) where the simple path is defined as a path without any loops. 
The intuition is that the presence of multiple paths between two entities indicates a stronger connection. For example, if an author has published more papers related to a concept $c_1$ compared to $c_2$,  we would assign a higher score to $c_1$ when retrieving concepts for that author.
We keep adding the score of $n_d$ when looping each starting entity $n_i$ since a target is more important if it shares connections with multiple source entities (\emph{line10}).
According to the score, we rank all reachable entities and return the top-K as target entities $\mathcal{N}_t$. We also select relations that start from entities in $\mathcal{N}_s$ to entities in $\mathcal{N}_t$ as target edges $\mathcal{E}_{s\rightarrow t}$. 

\subsection{A Dataflow System for Interactive Semantic Querying}


\subsubsection{System Requirements}
To achieve the semantic queries interactively and efficiently, 
we have incorporated a visual analytics system into our approach. 
The system's design was iteratively refined through weekly discussions with domain experts in the field of information retrieval and literature review. 
Based on this discussion, we have identified the following design tasks:

\begin{itemize}[leftmargin=*]\setlength\itemsep{-0.5em}
    \item \textbf{T1. Constructing interactive queries over SKG.} Querying SKG with structured queries, such as SPARQL, can have a high learning curve for users less experienced in programming. 
    To ensure a smooth and efficient knowledge discovery process,
    Our system should have an intuitive interface that allows users to construct semantic queries efficiently and interactively. 
    
    \item \textbf{T2. Visualizing heterogeneous data from SKG.} Due to the large volume and high diversity of SKG, data retrieved by the semantic queries can be difficult to grasp instantly. While visualization is a powerful tool to present patterns and generate insights, the system should provide clear and intuitive data visualization for queried data. 
    \item \textbf{T3. Building flexible pipelines for SKG.} Users with different backgrounds may have various goals for exploring SKG. The exploration process can involve multiple data queries, processing, and analysis to generate insight. 
    Our system should provide users with the ability to customize analysis pipelines.
    
    \item \textbf{T4. Accessing raw data in SKG.} While providing summarization of queried information is efficient for knowledge extraction, providing access to raw data is necessary to build user trust in the system and data. Also, our system should link the raw data to the web and provide an easy way for users to access it. 

\end{itemize}
\subsubsection{System Overview}

Guided by the tasks, we have designed and developed a dataflow system, as visualized in \autoref{fig:system}.
The system comprises a set of components with predefined inputs and outputs that users can choose and connect to construct their customized analysis pipeline (\textbf{T3}).
The component can be grouped into two categories: \emph{operators}  for data manipulation (\textbf{T1}) and \emph{viewers} for data visualization (\textbf{T2, T4}). 
The operators generate and manipulate data, which can be transmitted to downstream components of the system, while the visualizers present data meaningfully and offer insights without transmitting the data to the following components. 


\subsubsection{Operator Components}
 To facilitate effective exploration of the SKG, we have identified three critical stages for \autoref{alg:semanticquery}: (1) \emph{pre-execution}: finding the source entities $\mathcal{N}_s$, (2) \emph{execution}: retrieving target entities $\mathcal{N}_t$, and (3) \emph{post-execution}: analyzing the queried graph $\mathcal{E}_{s\rightarrow t}$.
 We abstract them into three sub-tasks, namely, node retrieval, node expansion, and graph comparison. To support each sub-task, we design three operators: \emph{querier, expander, and comparer.} 
 
 \textbf{Querier}
 allows users to retrieve entities using a sequence of keywords or keyphrases separated by commas (\autoref{fig:system}A). 
It performs a fuzzy match between user input and entity literal attributes, such as titles of papers, or full names of authors. 
The retrieved entities can serve as the source entities for \autoref{alg:semanticquery}. 

\textbf{Expander} plays a crucial role in efficiently exploring SKG by finding entities with a user-defined type.  
It is an implementation of \autoref{alg:semanticquery}.
To display the data (\textbf{T2}) and enable user interaction, we use a node-link diagram to depict source/target entities
as circles and their relationships as the connecting lines between them, as demonstrated in \autoref{fig:system}B.
During this process, multiple data can be used to satisfy various needs. Thus, we design three types of output for the \emph{expander} according to its subsequent components: 
\begin{itemize}[leftmargin=*]\setlength\itemsep{-0.5em}
    \item \emph{Target entities $\mathcal{N}_t$}: Once the \emph{expander} retrieves a list of target entities, they can be passed on to the next \emph{expander} as the next starting point for subsequent retrieval.
    
    \item \emph{Graph=($\mathcal{N}_s\cup\mathcal{N}_t, \mathcal{E}_{s\rightarrow t}$)}: 
    The graph in each \emph{expander} displays the edges from source to target entities. 
    The graph will be the output if connected to a \emph{comparer} for comparing one or multiple graphs.
    \item \emph{Graph=($\mathcal{N}_t, \mathcal{E}_{t\rightarrow t}$)}: Additionally, there are also connections among the target entities themselves. For instance, the citation relationships among papers or the collaborations between authors can be useful. It will pass this information when connecting to a \emph{visualizer}.
    \item \emph{Web URI}: Our SKG is integrated with DBpedia to expand knowledge. If users specify \emph{Concept} as the target entity type, they can click each \emph{Concept} node from \emph{expander} to transfer its DBpedia URL to a \emph{node viewer} for additional explanations and explorations. 
\end{itemize}


\textbf{Comparer} enables users to compare the graphs from 
two or more \emph{Expanders} and visualizes the merged graph with common nodes highlighted, as shown in \autoref{fig:system}C. 

\subsubsection{Viewer Components}
The viewers consist of three components that enable efficient access and visualization of various data. 

\textbf{Node Visualizer} is a highly effective tool for presenting data patterns to enable insight discovery, as shown in  \autoref{fig:system}F. 
During the node expansion step, the relationships between target entities $\mathcal{E}_{t\rightarrow t}$ can be informative in different usage scenarios.
To support the various needs, we incorporate two visualization charts that are the most effective in visualizing the graph structure.
The node-link diagram is efficient in showing the relationships among entities, such as citation networks between \emph{Paper} entities, or collaboration networks among \emph{Author} entities.
Another chart, the Sankey diagram, can provide a clear and organized visualization of networks among concepts. 
By positioning nodes along an axis, it can display concepts with different labels clearly. 


\textbf{Table Viewer} presents raw data in a tabular manner (\textbf{T4}). It enables users to develop confidence in our dataflow system by providing them with raw data from operator components (\autoref{fig:system}D).  

\textbf{Node Viewer} serves as a bridge between our SKG with DBpedia, displaying the corresponding DBpedia webpage when a user clicks on a \emph{Concept} of interest in \emph{Expander}, as shown in \autoref{fig:system}E.   
Additionally, the \emph{Web Viewer} offers a GUI interface for users to query (\autoref{fig:system}$e_1$) or perform SPARQL queries on the DBpedia (\autoref{fig:system}$e_2$).

\section{Evaluation of Concept Entity Extraction}

The quality of SKG heavily depends on the concept entities, which are extracted by our semi-supervised module. We conduct an evaluation of this process
from two perspectives: (1) whether the unsupervised pipeline can reduce labor work for users and (2) whether the fine-tuned NER model can accurately extract entities.
\begin{figure}[htp]
    \centering
    \includegraphics[width=\linewidth]{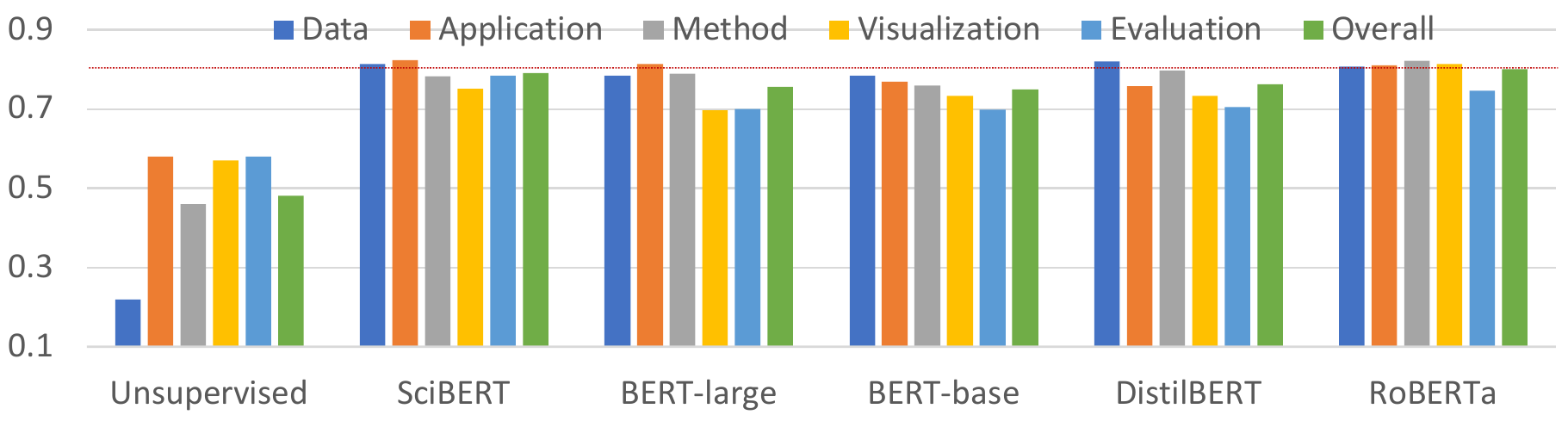}
    \caption{The F1 scores for five fine-tuned language models and an unsupervised pipeline on the Named Entity Recognition (NER) task.}
    \label{fig:eval}
    \vspace{-5mm}
\end{figure}

To assess this, we computed the f1-score of NER results by comparing the unsupervised results and the predicted results with the human-labeled ground truth. Traditionally, an exactly matched entity is considered one true positive case. However, we also perform entity normalization, which can compensate for some partially-matched entities. Therefore, we computed the f1-score at the token level instead of the entity level, which is more appropriate in our usage scenario. Our evaluation results are presented in \autoref{fig:eval}.

The unsupervised pipeline achieved an f1-score of over 0.5, except for the \emph{data} and \emph{method} classes. It proves the necessity of our first stage in reducing human labeling efforts. During the human annotation process, we identified some phenomena that could explain the low f1-score for the aforementioned two classes.
For the \emph{data} class, we found many descriptions such as ``visualization of [MASK] data''. Due to contextual information, the [MASK] data has a higher probability of answering the visualization question rather than the data question. However, during the annotation process, we converted those labels back to \emph{data}, bringing the low f1-score of \emph{data} class. 
For the \emph{method} class, the names of methods can be quite different and diverse, and the pre-defined rules fail to capture many methods, resulting in disagreement with human labeling.

We utilized an early-stop mechanism to fully fine-tune five language models and compared their performance. 
The results are shown in \autoref{fig:eval}. 
It is clear that all five models achieved good f1 scores compared to the unsupervised ones. 
Among those, the average f1-score of RoBERTa is the highest, and SciBERT is the second. We infer the reason is that RoBERTa has employed a dynamic masking mechanism for each batch during training that improves its generalizability on downstream tasks. 
On the other hand,  SciBERT's training on a large corpus of scientific text may explain its strong performance in our scenarios. 
Given RoBERTa's superior performance, we selected it for the case studies. 
Further details about model training and inference can be found in the supplemental material.

        

\section{Case study}
In this section, we present two real-world scenarios to (1) qualitatively and quantitatively evaluate the quality of SKG; (2) showcase the versatile nature of the dataflow system to perform semantic queries. 
The terms ``\emph{Concept} type'' and ``dimension'' are used interchangeably throughout the cases. 
\vspace{-2mm}
\subsection{Information Summarization of Trust in ML}

\begin{figure}
    \centering
    \includegraphics[width=\linewidth]{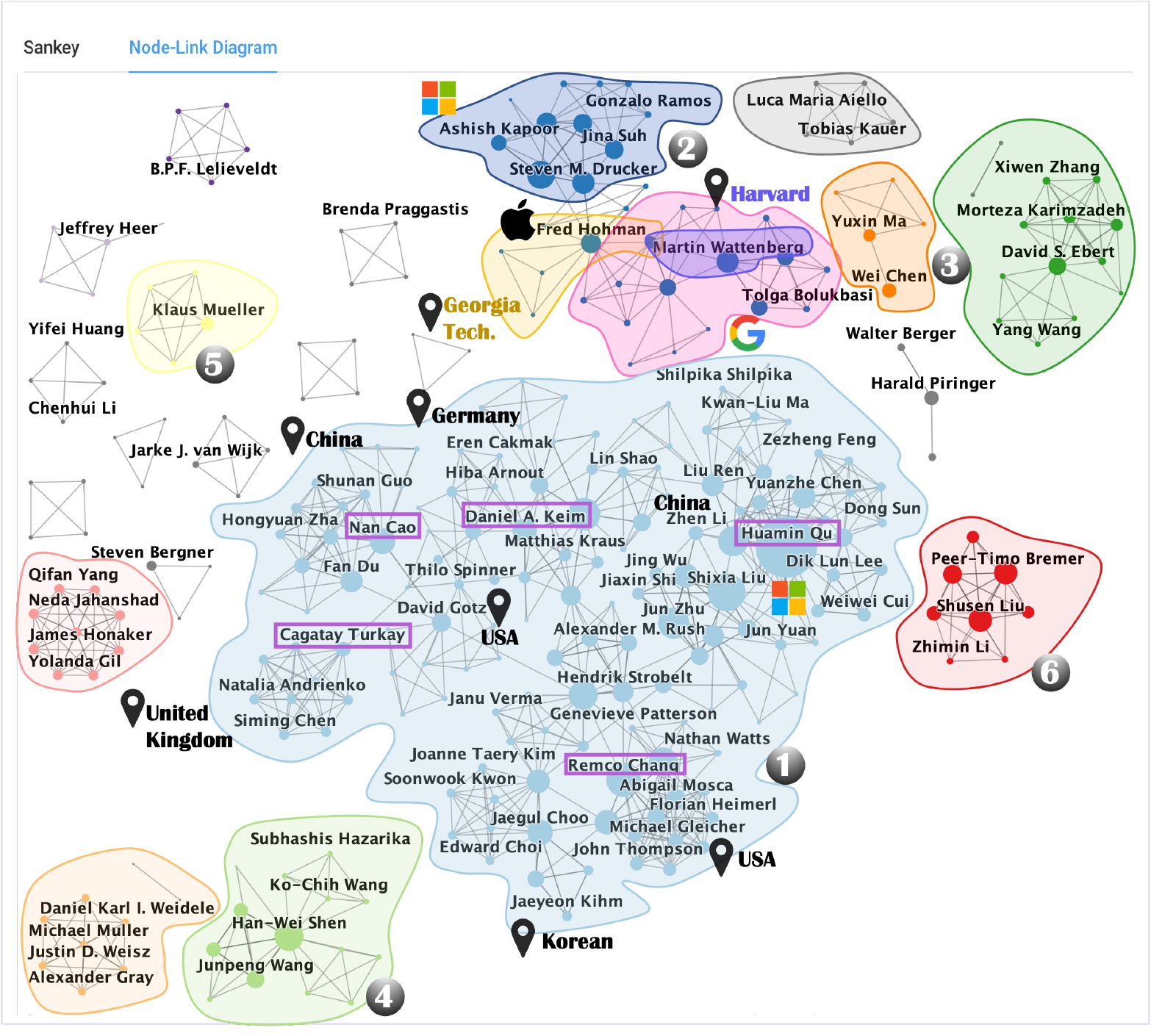}
    \caption{The collaboration network of scholars on the topic of improving trust in machine learning through visualization.}
    \label{fig:case_ml}
    \vspace{-4mm}
\end{figure}

The purpose of this study is to demonstrate the efficiency of SKG in summarizing a set of target papers from multiple perspectives. 
Specifically, we aim to use our dataflow system to follow a STAR paper published in EuroVis 2020 that focuses on improving trust in machine learning models through visualizations~\cite{chatzimparmpas2020state}. Our focus is on evaluating (1) whether our SKG covers enough meta-data with high-quality to support literature analysis in the visualization domain and (2) whether our SKG 
contains useful semantic information to automate the summarization, reducing human efforts to manually collect, read, label and summarize papers.  

There are 200 surveyed papers included in the original STAR. 
To obtain the paper list, we crawled their website\footnote{\url{https://trustmlvis.lnu.se/}} and excluded papers published after 2020 when the STAR was published.
We used these paper titles as input and utilized a \emph{querier} to retrieve papers from SKG, resulting in 162 out of 200 papers retrieved. 
This confirms that our SKG covers the main conferences/journals in the VIS literature, providing robust support for further study.

\subsubsection{Co-authorship Analysis}
Given the retrieved papers, building the collaboration network of authors requires certain data processing steps. 
Our SKG, along with the dataflow system, can be used to facilitate this process.
By using an \emph{expander}, \emph{Author} nodes can be retrieved easily and a \emph{visualizer} can be connected to display the relationships among authors.
\autoref{fig:case_ml} displays the network view.
Larger nodes represent authors with more connections, and we have hidden some labels to avoid visual clutter.

From the figure, we can easily confirm the same significant findings as in the STAR paper. For example, we can identify a large collaboration network in the middle where we can spot some top scholars \circled{1}, including Huamin Qu, Remco Chang, Daniel A. Keim, and Cagatay Turday. Additionally, from the top of \autoref{fig:case_ml}, we can see clusters  related to industrial efforts towards ML trust \circled{2}, such as Steven M. Drucker and Tolga Bolukbasi. The STAR paper also mentions smaller clusters of collaborations, which we can also identify from \autoref{fig:case_ml}, such as David S. Ebert \circled{3}, Wei Chen \circled{3}, Klaus Mueller \circled{5}, Han-Wei Shen \circled{4}, and Valerio Pascucci  \circled{6}.

In addition to the findings in the original STAR paper, we highlight some additional collaboration patterns. Collaboration is quite diverse between (1) academia and industry. For example, companies such as Google, Microsoft, and Apple have close collaborations with academic institutions such as Georgia Tech and Harvard \circled{2}. Additionally, Microsoft Asia collaborates with many top universities in China \circled{1}. (2) various universities across different countries and regions, including the USA, Korea, the United Kingdom, Germany, and China, form a large collaboration network \circled{1}. Given the development of machine learning, we believe that such diverse collaboration is critical for applying visualization to build trust in machine learning.

Through these explorations, we affirm that the papers contained in the SKG are extensive to support a literature review. Also, the meta-data, such as authors, are of high quality to provide insights for meta-data analysis. 

\subsubsection{Topic Analysis}
\renewcommand{\arraystretch}{0.5} 

\begin{table}[]
\small
\caption{Comparison of automatically summarized concepts in SKG with the expert summarization in STAR paper.}
\vspace{-3mm}
\begin{tabular}{llll}
\hline
\rowcolor[HTML]{EFEFEF} 
\textbf{\begin{tabular}[c]{@{}l@{}}Domain \\ (88.9\%)\end{tabular}} &
  \textbf{\begin{tabular}[c]{@{}l@{}}ML Types\\ (100\%)\end{tabular}} &
  \textbf{\begin{tabular}[c]{@{}l@{}}ML Methods\\ (81.8\%)\end{tabular}} &
  \textbf{\begin{tabular}[c]{@{}l@{}}Visual \\ Representation \\ (82.3\%)\end{tabular}} \\ \hline
\begin{tabular}[c]{@{}l@{}}Biology (3)\\ \\ \\ \\ Business (1)\\ \\ \\ \\ Computer \\ Vision (5)\\ \\ \\ \\ Computer (18)\\ \\ \\ \\ Health (8)\\ \\ \\ \\ \\ Humanities(10)\\ \\ \\ \\ \\ Nutrition (0)\\ \\ \\ \\ \\ Simulation (8)\\ \\ \\ Social/(3)\\ Socioeconomic\end{tabular} &
  \begin{tabular}[c]{@{}l@{}}Classification(45)\\ \\ \\ \\ Regression(9)\\ \\ \\ \\ Association(1)\\ \\ \\ \\ \\ Clustering(167)\\ \\ \\ \\ Dimensionality\\ Reduction (31)\\ \\ \\ \\ Supervised (5)\\ \\ \\ \\ \\ Semi-supervised (1)\\ \\ \\ \\ \\ Reinforcement (3)\\ \\ \\ \\ Unsupervised (1)\end{tabular} &
  \begin{tabular}[c]{@{}l@{}}Convolutional   \\ Neural Network (8)\\ \\ Deep  Convolutional \\ Network (0)\\ \\ Deep Feed \\ Forward (0)\\ \\ \\ Deep Neural \\ Network (5)\\ \\ Deep Q-Network (5)\\ \\ Generative \\ Adversarial \\ Network (10)\\ \\ \\ Long Short-\\ Term Memory (2)\\ \\ Recurrent Neural \\ Network (7)\\ \\ \\ Variational \\ Auto-Encoder (1)\\ \\ \\ Dimensionality\\  Reduction (31)\\ \\ \\ Ensemble \\ Learning (7)\end{tabular} &
  \begin{tabular}[c]{@{}l@{}}Bar Chart (1)\\ \\ Box Plot (0)\\ \\ Matrix (15)\\ \\ Glyph/Icon\\ /Thumbnail (4)\\ \\ Grid-based (3)\\ \\ Heatmap (2)\\ \\ Histogram (1)\\ \\ Icicle Plot (1)\\ \\ Line Chart (0)\\ \\ Node-link \\ Diagram (2)\\ \\ Parallel \\ Coordinate (7)\\ \\ Pixel-based (0)\\ \\ Radial Layout (1)\\ \\ Scatterplot (47)\\ \\ Similarity \\ Layout (0)\\ \\ Table/List (14)\\ \\ Treemap (2)\end{tabular} \\ \hline
\end{tabular}
\label{tab:case_ml}
\vspace{-6mm}
\end{table}

This study aims to evaluate the quality of  extracted \emph{Concepts} in SKG by comparing them with concepts summarized by domain experts. The result will illustrate how SKG can help in the semantic summarization of papers. 
In the STAR, the experts summarize the topics from multiple perspectives. We chose four dimensions that are relevant to the relationships between \emph{Concepts} and \emph{Papers} in SKG. 
Specifically, four dimensions include \emph{domain, ML types, ML Methods, and visual representations}, which are corresponded to \emph{Concepts} with semantic roles of \emph{application, task, method, visualization} in SKG.

To perform the comparison, we began by retrieving all the \emph{Concept} entities from 162 \emph{Paper} entities from SKG, denoted as $\mathcal{C}$. 
From the STAR, we also obtained a set of concepts summarized by the domain experts as the ground truth,  referred to as $\mathcal{C'}$. 
For each concept $c' \in \mathcal{C'}$, we count its occurrence that appeared in $\mathcal{C}$. We list all the ground truth concepts $c'$ with their occurrence in \autoref{tab:case_ml}.
Also, in the header, the set coverage of each dimension is reported. 
The overall coverage is over 80\%, indicating that our SKG can capture information that is considered as important by the domain experts. 
However, we noticed that the number of papers we found is smaller than the number reported in the STAR. The reason is that some concepts are not explicitly mentioned in the abstract paragraph due to the limited length, which leaves room for further data integration in SKG. 
In addition, we observed that for each ground truth \emph{Concept} entity $c' \in \mathcal{C'}$, we were able to locate a range of more specific concepts in $C$. For instance, when examining the concept of ``classification'', we identified ``binary classification,'' ``multi-label classification,''  ``sentiment analysis'' and more.
We believe this rich data evidence can inspire researchers to refine their summarization strategies, accelerating the 
the knowledge discovery process and making it easier for them to perform  literature reviews.

\subsection{Knowledge Discovery of Text Mining}
\begin{figure*}
    \centering
    \includegraphics[width=\textwidth]{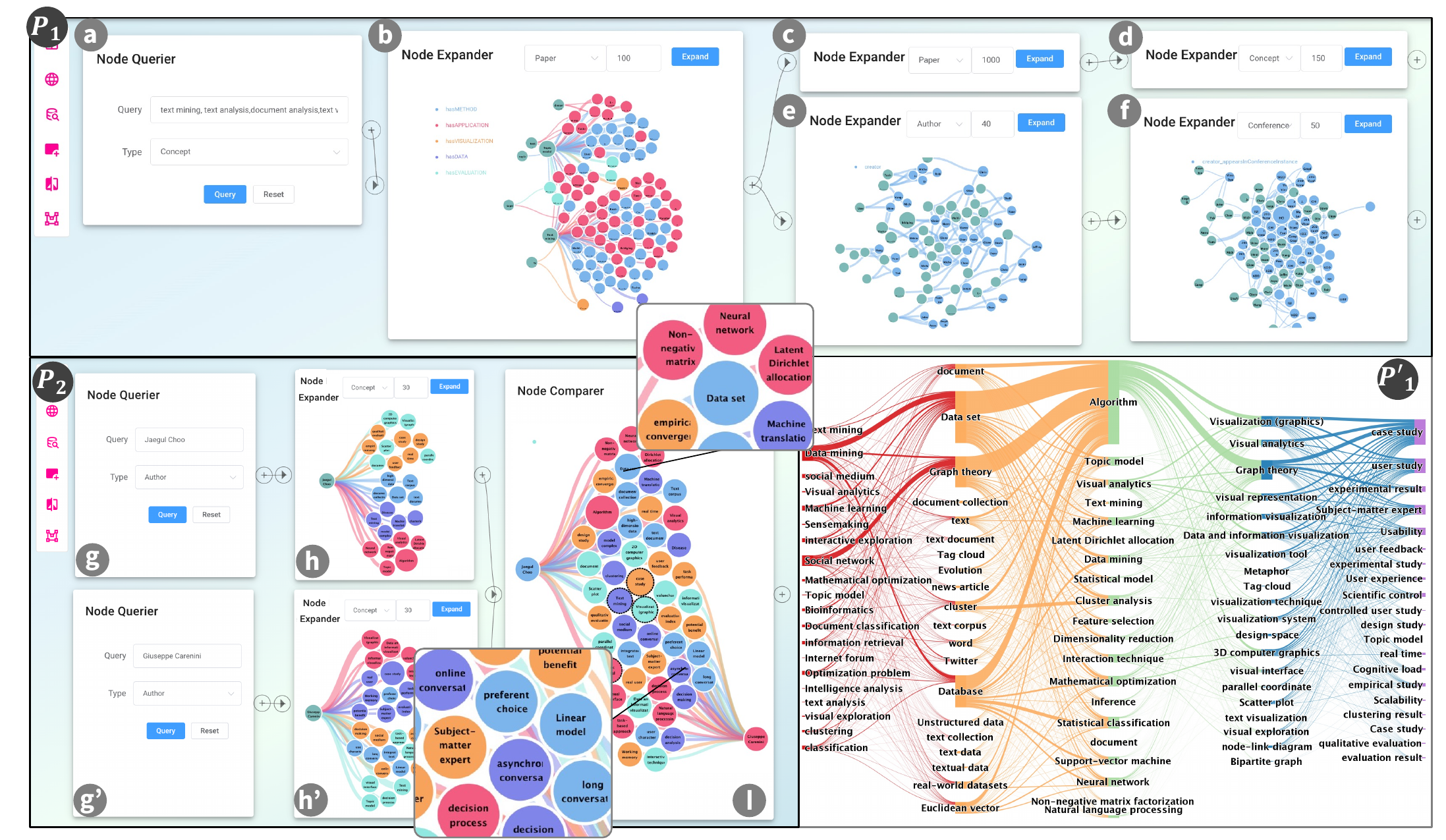}
    \caption{Exploring knowledge discovery through text mining using two pipelines: (P1) document retrieval and summarization; (P2) comparing research interests between \emph{Jaegul Choo} and \emph{Giuseppe Carenini}; (P1') a Sankey diagram generated by a \emph{visualizer} after component d in P1.}
    \label{fig:case_tm}
    \vspace{-6mm}
\end{figure*}

With the wealth of information available in SKG, we illustrate how the dataflow system can effectively support interactive queries for various tasks. 
In this case, we invite a Ph.D. student, Emily, with a focus on  \emph{text analysis and visualization}, to use our dataflow system for her information-seeking needs related to her research interests.


\subsubsection{Automatic Literature Review}
Emily intended to perform an automatic literature review with the help of SKG. 
First, she was wondering if she could combine multi-type of information in SKG to perform document retrieval since single keyword matching might miss some related papers. 
 To achieve it, Emily decided first to select some \emph{papers} that are connected to target \emph{concepts}, and then find more relevant papers by utilizing citation relationships.  
 The analysis pipeline is shown in \autoref{fig:case_tm} ($P_1:\circled{a}\rightarrow \circled{b} \rightarrow \circled{c}\rightarrow \circled{d}$).
 Emily began with an \emph{querier} (\circled{a}) and typed in some phrases to retrieve relevant \emph{concepts}, and the phrases used include ``text mining'', ``text analysis'', ``document analysis'', ``text visualization'', ``topic model'', ``topic analysis''. 
After retrieving \emph{concept} from SKG, Emily performed document retrieval with two consecutive \emph{expanders}.
The first one (\circled{b}) utilized the relationships from \emph{concepts} to \emph{papers}, while the second one used the citation relationships (\circled{c}). 
Emily set the number of returned and obtained $1000$ papers in total. 
She is satisfied that various relationship types in SKG can be utilized to retrieve documents in succinct way without the troublesome of relying on multiple resources. 

Later, Emily is curious about the quality of the retrieved papers and wondering whether the \emph{concepts} in SKG can accelerate the evaluation process. 
To do so, she utilized another \emph{expander} (\circled{d}) to obtain the top 150 \emph{concepts} connected to previous retrieved papers.
Upon examining the \emph{application} dimension,
Emily found several tasks that are important in this literature~\cite{liu2018bridging}, such as \emph{information retrieval, clustering, classification, visual exploration, and social network}. 
This dimension also included several concepts that described  domain information, such as \emph{bioinformatics, machine learning, sensemaking}. 
Later, Emily started to check the \emph{visualization} dimension 
and found specific charts used for text visualization,  including 
\emph{scatterplot}, \emph{node-link diagram and bipartite graph}. 
When hovering the concept of interest, the connected nodes are highlighted, making it clear to locate specific papers for further reading. 
Next, Emily began the exploration of \emph{method} dimension.
she found various model types, described as follows:
\begin{itemize}[leftmargin=*]\setlength\itemsep{-0.5em}
    \item \emph{for classification}: Support Vector Machine, statistical classification; 
    \item \emph{for clustering}: clustering analysis, dimensionality reduction;
    \item \emph{for inference}: inference, statistical model;
    \item \emph{topic models}: topic model, Latent Dirichlet Allocation, Non-negative Matrix Factorization;
    \item \emph{language models}: neural networks, natural language processing.
\end{itemize} Through the exploration, Emily is surprised that the automatic summarization well aligns with her domain knowledge. This confirmed her belief that the concept extraction step during SKG construction is quite important as it saves human efforts by providing a  concrete and accurate semantic summary before they spend time reading in detail. 

To gain a better understanding of the relationships between these \emph{Concept} entities, Emily also loaded a \emph{visualizer} after \circled{d} to create a Sankey diagram, as shown in \autoref{fig:case_tm} ($P'_1$). 
The links in the diagram were formed based on 
the co-occurrence of two \emph{Concept} entities
in the SKG. 
Emily believes that this information can be highly beneficial, as it  provides researchers with valuable empirical evidence that can aid in making 
informed decisions and 
generating precise summaries. For instance, the relationships between the application dimension and the method dimension can provide insights into which text-mining methods are proposed for specific tasks or applications.

In this case, the relationships between \emph{Paper} and \emph{Concept} in SKG are efficiently leveraged in both retrieval and summarization stages. Moreover, the diverse  semantic meanings embedded within these relationships offer a summary from multiple perspectives.

\subsubsection{Comparative Study of Scholarly Profiling}
Emily also would like to evaluate whether the SKG can profile users from multiple perspectives. Thus, based on the retrieved papers, Emily further extracted the top 40  authors with the most connections, as shown in \autoref{fig:case_tm} ($P_1: \circled{b}\rightarrow\circled{e}$).

She intended to profile users on the publication conferences of their research papers. 
Because publications not only provide insights into the focus of researchers' interests but also offers an indication of the impact of their research. Thus, she further expanded the graph with  \emph{conference} nodes (\circled{f}). 
By hovering one node, all the related nodes will be highlighted. Enabled by this feature, Emily 
hovered each \emph{conference} node and identified two popular conferences with the most publications: IEEE Transactions on Visualization and Computer Graphics
 and Comput. Graph. Forum. 
Also, she hovered 
each \emph{author} node to profile individual scholars based on their conference affiliations. 
For example, she noted that \emph{Giuseppe Carenini} had more publications on CHI (\emph{Human-Computer Interaction}) and IUI (\emph{Intelligent User Interfaces}) while Jaegul Choo focused more on \emph{TVCG}, which raised questions about potential research differences between the two scholars. 

To analyze the research interests of the two authors, Emily built another analysis pipeline to perform a comparative analysis (\autoref{fig:case_tm}$P_2$).
She started by using a \emph{querier} to search for \emph{Jaegual Choo} and connected an \emph{expander} to find his research topics, as shown in \circled{g}$\rightarrow$\circled{h}. 
She performed the same operation for \emph{Giuseppe Carenini} and obtained his research topics in \circled{g'}$\rightarrow$\circled{h'}.
Emily then utilized a \emph{comparer} to perform a graph comparison between the two authors (\circled{I}). 

From the result (\circled{I}),
Emily discovered that \emph{Giuseppe Carenini}'s research interests were focused  on ``Natural Language Processing,'' specifically on conversational topics, as she found ``online conversation'', ``asynchronous  conversations'', and ``long conversations''. 
Additionally, she noticed concepts related to  ``decision analysis,'' and ``interaction technique,'' which led her to hypothesize that his research topic was related to building visual interactive systems to assist the decision-making process related to conversations. Later, Emily confirmed her hypothesis by finding the related papers~\cite{hoque2014convis, hoque2015convisit}.
For \emph{Jaegul Choo}, Emily identified several algorithms related to topic modeling, such as ``Non-negative Matrix Factorization,'' ``Latent Dirichlet Allocation,'' and ``machine translation''. 
She also noted several visualization charts, such as ``parallel coordinate,'' and ``scatter plot''. 
Based on these findings, she inferred that his research interests might be diverse, spanning multiple topics.
To gain a better understanding of
\emph{Jaegul Choo}'s research, Emily conducted an online search and discovered that he had published several papers on Non-negative Matrix Factorization ~\cite{choo2013utopian, kuang2015nonnegative} and image-to-image translation~\cite{kuang2015nonnegative, cho2019image}, which support her findings in SKG.
From these explorations, Emily confirmed that querying over SKG is more efficient than conducting an online search and scanning through all the publications to generate insights.

In this case, SKG integrates various meta-information and  semantic information to support author profiling. 
Without such integration, users have to conduct online searches from multiple data sources, keep a record of them, and manually integrate the information, which could hinder the efficiency of the information-seeking process. 


\vspace{-2mm}
\section{Discussion and Future Work}

In this section, we discuss several major advantages of SKG.

\textbf{Extensibility}: Extensibility is an important property of the knowledge graph, which means that our SKG can easily incorporate new knowledge from upcoming publications and  expand with additional information, such as affiliations or domain experts defined knowledge. Furthermore, the SKG can be seamlessly integrated with other knowledge graphs to enhance literature exploration and analysis.

\textbf{Usability}: The usability of SKG extends beyond information retrieval and analysis, as it can be utilized in various downstream tasks. For instance, it can power recommendation systems to generate personalized suggestions and support natural language queries. Moreover, SKG can serve as input for Graph Neural Networks, enabling node classification, link prediction, and community detection. We believe this domain-specific dataset can benefit the research community in numerous ways.

\textbf{Transferability}:
Although our SKG was built from documents in the visualization domain for demonstration purposes, our approach can be transferred to other domains with minimal modifications, only requiring the definition of concepts in the relevant literature. Furthermore, our trained NER model can be readily deployed to identify and classify entities in papers of other domains, which can prove useful in various other scenarios beyond knowledge graph construction.

With the explosion of data online, the knowledge graph is a trend as it integrates various data sources into a unified representation. This data integration facilitates efficient data utilization later. 
Going forward, we anticipate the continued growth and integration of new knowledge into visualization SKGs. 
Besides,  the heterogeneous information captured in the SKG presents an exciting opportunity to explore the insights that could be gained from machine learning. 
It would be interesting to see how machine learning algorithms could benefit domain experts by uncovering previously unknown relationships and patterns in the data. 
\section{conclusion}
In this work, we propose utilizing a Semantic Knowledge Graph (SKG) to improve information retrieval and analysis in academic literature. To construct SKG, we introduce a semi-supervised module that extracts and categorizes knowledge from unstructured text.
We then showcase the utility of SKG in academic papers exploration by summarizing the supported semantic queries and implementing a dataflow system to enable interactive and efficient querying.
Our approach is demonstrated in the field of visualization literature through multiple real-world case studies that effectively demonstrate the usefulness of the approach.

\bibliographystyle{abbrv-doi}
\bibliography{template}

\begin{thebibliography}{10}

\bibitem{al2020end}
K.~Al-Khatib, Y.~Hou, H.~Wachsmuth, C.~Jochim, F.~Bonin, and B.~Stein.
\newblock End-to-end argumentation knowledge graph construction.
\newblock In {\em Proceedings of the AAAI conference on artificial
  intelligence}, vol.~34, pp. 7367--7374, 2020.

\bibitem{allam2012question}
A.~M.~N. Allam and M.~H. Haggag.
\newblock The question answering systems: A survey.
\newblock {\em International Journal of Research and Reviews in Information
  Sciences (IJRRIS)}, 2(3), 2012.

\bibitem{angeli2015leveraging}
G.~Angeli, M.~J.~J. Premkumar, and C.~D. Manning.
\newblock Leveraging linguistic structure for open domain information
  extraction.
\newblock In {\em Proceedings of the 53rd Annual Meeting of the Association for
  Computational Linguistics and the 7th International Joint Conference on
  Natural Language Processing (Volume 1: Long Papers)}, pp. 344--354, 2015.

\bibitem{auer2007dbpedia}
S.~Auer, C.~Bizer, G.~Kobilarov, J.~Lehmann, R.~Cyganiak, and Z.~Ives.
\newblock Dbpedia: A nucleus for a web of open data.
\newblock In {\em The Semantic Web: 6th International Semantic Web Conference,
  2nd Asian Semantic Web Conference, ISWC 2007+ ASWC 2007, Busan, Korea,
  November 11-15, 2007. Proceedings}, pp. 722--735. Springer, 2007.

\bibitem{berners2001semantic}
T.~Berners-Lee, J.~Hendler, and O.~Lassila.
\newblock The semantic web.
\newblock {\em Scientific american}, 284(5):34--43, 2001.

\bibitem{bollacker2008freebase}
K.~Bollacker, C.~Evans, P.~Paritosh, T.~Sturge, and J.~Taylor.
\newblock Freebase: a collaboratively created graph database for structuring
  human knowledge.
\newblock In {\em Proceedings of the 2008 ACM SIGMOD international conference
  on Management of data}, pp. 1247--1250, 2008.

\bibitem{buscaldi2019mining}
D.~Buscaldi, D.~Dess{\`\i}, E.~Motta, F.~Osborne, and D.~Reforgiato~Recupero.
\newblock Mining scholarly publications for scientific knowledge graph
  construction.
\newblock In {\em The Semantic Web: ESWC 2019 Satellite Events: ESWC 2019
  Satellite Events, Portoro{\v{z}}, Slovenia, June 2--6, 2019, Revised Selected
  Papers 16}, pp. 8--12. Springer, 2019.

\bibitem{chansanam2022culture}
W.~Chansanam, Y.~Jaroenruen, N.~Kaewboonma, and T.~Kulthida.
\newblock Culture knowledge graph construction techniques.
\newblock {\em Education for Information}, (Preprint):1--32, 2022.

\bibitem{chatzimparmpas2020state}
A.~Chatzimparmpas, R.~M. Martins, I.~Jusufi, K.~Kucher, F.~Rossi, and
  A.~Kerren.
\newblock The state of the art in enhancing trust in machine learning models
  with the use of visualizations.
\newblock In {\em Computer Graphics Forum}, vol.~39, pp. 713--756. Wiley Online
  Library, 2020.

\bibitem{chen2019automatic}
H.~Chen and X.~Luo.
\newblock An automatic literature knowledge graph and reasoning network
  modeling framework based on ontology and natural language processing.
\newblock {\em Advanced Engineering Informatics}, 42:100959, 2019.

\bibitem{cho2019image}
W.~Cho, S.~Choi, D.~K. Park, I.~Shin, and J.~Choo.
\newblock Image-to-image translation via group-wise deep whitening-and-coloring
  transformation.
\newblock In {\em Proceedings of the IEEE/CVF Conference on Computer Vision and
  Pattern Recognition}, pp. 10639--10647, 2019.

\bibitem{choo2013utopian}
J.~Choo, C.~Lee, C.~K. Reddy, and H.~Park.
\newblock Utopian: User-driven topic modeling based on interactive nonnegative
  matrix factorization.
\newblock {\em IEEE transactions on visualization and computer graphics},
  19(12):1992--2001, 2013.

\bibitem{deagen2022fair}
M.~E. Deagen, J.~P. McCusker, T.~Fateye, S.~Stouffer, L.~C. Brinson, D.~L.
  McGuinness, and L.~S. Schadler.
\newblock Fair and interactive data graphics from a scientific knowledge graph.
\newblock {\em Scientific Data}, 9(1):1--11, 2022.

\bibitem{dessi2021generating}
D.~Dess{\`\i}, F.~Osborne, D.~R. Recupero, D.~Buscaldi, and E.~Motta.
\newblock Generating knowledge graphs by employing natural language processing
  and machine learning techniques within the scholarly domain.
\newblock {\em Future Generation Computer Systems}, 116:253--264, 2021.

\bibitem{dessi2020ai}
D.~Dess{\`\i}, F.~Osborne, D.~Reforgiato~Recupero, D.~Buscaldi, E.~Motta, and
  H.~Sack.
\newblock Ai-kg: an automatically generated knowledge graph of artificial
  intelligence.
\newblock In {\em The Semantic Web--ISWC 2020: 19th International Semantic Web
  Conference, Athens, Greece, November 2--6, 2020, Proceedings, Part II 19},
  pp. 127--143. Springer, 2020.

\bibitem{ding2022tell}
J.~Ding, T.~Xiang, Z.~Ou, W.~Zuo, R.~Zhao, C.~Lin, Y.~Zheng, and B.~Liu.
\newblock Tell me how to survey: Literature review made simple with automatic
  reading path generation.
\newblock In {\em 2022 IEEE 38th International Conference on Data Engineering
  (ICDE)}, pp. 3426--3438. IEEE, 2022.

\bibitem{e2015fedviz}
S.~S. e~Zainab, M.~Saleem, Q.~Mehmood, D.~Zehra, S.~Decker, and A.~Hasnain.
\newblock Fedviz: A visual interface for sparql queries formulation and
  execution.
\newblock In {\em VOILA@ ISWC}, p.~49, 2015.

\bibitem{feng2021small}
D.~Feng and H.~Chen.
\newblock A small samples training framework for deep learning-based automatic
  information extraction: Case study of construction accident news reports
  analysis.
\newblock {\em Advanced Engineering Informatics}, 47:101256, 2021.

\bibitem{gomez2018visualizing}
J.~G{\'o}mez-Romero, M.~Molina-Solana, A.~Oehmichen, and Y.~Guo.
\newblock Visualizing large knowledge graphs: A performance analysis.
\newblock {\em Future Generation Computer Systems}, 89:224--238, 2018.

\bibitem{guo2022automatic}
L.~Guo, F.~Yan, T.~Li, T.~Yang, and Y.~Lu.
\newblock An automatic method for constructing machining process knowledge base
  from knowledge graph.
\newblock {\em Robotics and Computer-Integrated Manufacturing}, 73:102222,
  2022.

\bibitem{he2019aloha}
X.~He, R.~Zhang, R.~Rizvi, J.~Vasilakes, X.~Yang, Y.~Guo, Z.~He, M.~Prosperi,
  J.~Huo, J.~Alpert, et~al.
\newblock Aloha: developing an interactive graph-based visualization for
  dietary supplement knowledge graph through user-centered design.
\newblock {\em BMC medical informatics and decision making}, 19(4):1--18, 2019.

\bibitem{hirsch2009interactive}
C.~Hirsch, J.~Hosking, and J.~Grundy.
\newblock Interactive visualization tools for exploring the semantic graph of
  large knowledge spaces.
\newblock In {\em Workshop on Visual Interfaces to the Social and the Semantic
  Web (VISSW2009)}, vol. 443, pp. 11--16. Citeseer, 2009.

\bibitem{hoque2014convis}
E.~Hoque and G.~Carenini.
\newblock Convis: A visual text analytic system for exploring blog
  conversations.
\newblock In {\em Computer Graphics Forum}, vol.~33, pp. 221--230. Wiley Online
  Library, 2014.

\bibitem{hoque2015convisit}
E.~Hoque and G.~Carenini.
\newblock Convisit: Interactive topic modeling for exploring asynchronous
  online conversations.
\newblock In {\em Proceedings of the 20th International Conference on
  Intelligent User Interfaces}, pp. 169--180, 2015.

\bibitem{isenberg2016visualization}
P.~Isenberg, T.~Isenberg, M.~Sedlmair, J.~Chen, and T.~M{\"o}ller.
\newblock Visualization as seen through its research paper keywords.
\newblock {\em IEEE Transactions on Visualization and Computer Graphics},
  23(1):771--780, 2016.

\bibitem{jelodar2019latent}
H.~Jelodar, Y.~Wang, C.~Yuan, X.~Feng, X.~Jiang, Y.~Li, and L.~Zhao.
\newblock Latent dirichlet allocation (lda) and topic modeling: models,
  applications, a survey.
\newblock {\em Multimedia Tools and Applications}, 78:15169--15211, 2019.

\bibitem{9239940}
B.~Jiang, X.~You, K.~Li, T.~Li, X.~Zhou, and L.~Tan.
\newblock Interactive analysis of epidemic situations based on a spatiotemporal
  information knowledge graph of covid-19.
\newblock {\em IEEE Access}, 10:46782--46795, 2022. doi: {{%
10\hspace{.1pt}\discretionary{.}{%
}{.}\hspace{.4pt}1109\discretionary{/}{%
}{/}ACCESS\hspace{.1pt}\discretionary{.}{%
}{.}\hspace{.4pt}2020\hspace{.1pt}\discretionary{.}{%
}{.}\hspace{.4pt}3033997}}


\bibitem{kuang2015nonnegative}
D.~Kuang, J.~Choo, and H.~Park.
\newblock Nonnegative matrix factorization for interactive topic modeling and
  document clustering.
\newblock {\em Partitional clustering algorithms}, pp. 215--243, 2015.

\bibitem{liu2018bridging}
S.~Liu, X.~Wang, C.~Collins, W.~Dou, F.~Ouyang, M.~El-Assady, L.~Jiang, and
  D.~A. Keim.
\newblock Bridging text visualization and mining: A task-driven survey.
\newblock {\em IEEE transactions on visualization and computer graphics},
  25(7):2482--2504, 2018.

\bibitem{lo-wang-2020-s2orc}
K.~Lo, L.~L. Wang, M.~Neumann, R.~Kinney, and D.~Weld.
\newblock {S}2{ORC}: The semantic scholar open research corpus.
\newblock In {\em Proceedings of the 58th Annual Meeting of the Association for
  Computational Linguistics}, pp. 4969--4983. Association for Computational
  Linguistics, Online, July 2020. doi: {{%
10\hspace{.1pt}\discretionary{.}{%
}{.}\hspace{.4pt}18653\discretionary{/}{%
}{/}v1\discretionary{/}{%
}{/}2020\hspace{.1pt}\discretionary{.}{%
}{.}\hspace{.4pt}acl\discretionary{%
}{-}{-}main\hspace{.1pt}\discretionary{.}{%
}{.}\hspace{.4pt}447}}


\bibitem{luan2018multi}
Y.~Luan, L.~He, M.~Ostendorf, and H.~Hajishirzi.
\newblock Multi-task identification of entities, relations, and coreference for
  scientific knowledge graph construction.
\newblock {\em arXiv preprint arXiv:1808.09602}, 2018.

\bibitem{maulud2021state}
D.~H. Maulud, S.~R. Zeebaree, K.~Jacksi, M.~A.~M. Sadeeq, and K.~H. Sharif.
\newblock State of art for semantic analysis of natural language processing.
\newblock {\em Qubahan Academic Journal}, 1(2):21--28, 2021.

\bibitem{mayfield2003information}
J.~Mayfield, T.~Finin, et~al.
\newblock Information retrieval on the semantic web: Integrating inference and
  retrieval.
\newblock In {\em Proceedings of the SIGIR Workshop on the Semantic Web}, 2003.

\bibitem{miller1995wordnet}
G.~A. Miller.
\newblock Wordnet: a lexical database for english.
\newblock {\em Communications of the ACM}, 38(11):39--41, 1995.

\bibitem{mondal2021end}
I.~Mondal, Y.~Hou, and C.~Jochim.
\newblock End-to-end nlp knowledge graph construction.
\newblock {\em arXiv preprint arXiv:2106.01167}, 2021.

\bibitem{moro2014entity}
A.~Moro, A.~Raganato, and R.~Navigli.
\newblock Entity linking meets word sense disambiguation: a unified approach.
\newblock {\em Transactions of the Association for Computational Linguistics},
  2:231--244, 2014.

\bibitem{murdock2015visualization}
J.~Murdock and C.~Allen.
\newblock Visualization techniques for topic model checking.
\newblock In {\em Proceedings of the AAAI Conference on Artificial
  Intelligence}, vol.~29, 2015.

\bibitem{9355442}
R.~Nararatwong, N.~Kertkeidkachorn, and R.~Ichise.
\newblock Knowledge graph visualization: Challenges, framework, and
  implementation.
\newblock In {\em 2020 IEEE Third International Conference on Artificial
  Intelligence and Knowledge Engineering (AIKE)}, pp. 174--178, 2020. doi: {{%
10\hspace{.1pt}\discretionary{.}{%
}{.}\hspace{.4pt}1109\discretionary{/}{%
}{/}AIKE48582\hspace{.1pt}\discretionary{.}{%
}{.}\hspace{.4pt}2020\hspace{.1pt}\discretionary{.}{%
}{.}\hspace{.4pt}00034}}


\bibitem{narechania2021vitality}
A.~Narechania, A.~Karduni, R.~Wesslen, and E.~Wall.
\newblock Vitality: Promoting serendipitous discovery of academic literature
  with transformers \& visual analytics.
\newblock {\em IEEE Transactions on Visualization and Computer Graphics},
  28(1):486--496, 2021.

\bibitem{navigli2012babelnet}
R.~Navigli and S.~P. Ponzetto.
\newblock Babelnet: The automatic construction, evaluation and application of a
  wide-coverage multilingual semantic network.
\newblock {\em Artificial intelligence}, 193:217--250, 2012.

\bibitem{pellegrino2020move}
M.~A. Pellegrino, V.~Scarano, and C.~Spagnuolo.
\newblock Move cultural heritage knowledge graphs in everyone’s pocket.
\newblock {\em Semantic Web}, (Preprint):1--37, 2020.

\bibitem{qiu2020tax}
Y.~Qiu, Y.~Qiao, S.~Yang, and J.~Yang.
\newblock Tax-kg: Taxation big data visualization system for knowledge graph.
\newblock In {\em 2020 IEEE 5th International Conference on Signal and Image
  Processing (ICSIP)}, pp. 425--429. IEEE, 2020.

\bibitem{2016arXiv160605250R}
P.~{Rajpurkar}, J.~{Zhang}, K.~{Lopyrev}, and P.~{Liang}.
\newblock {SQuAD: 100,000+ Questions for Machine Comprehension of Text}.
\newblock {\em arXiv e-prints}, p. arXiv:1606.05250, 2016.

\bibitem{salatino2019cso}
A.~A. Salatino, F.~Osborne, T.~Thanapalasingam, and E.~Motta.
\newblock The cso classifier: Ontology-driven detection of research topics in
  scholarly articles.
\newblock In {\em Digital Libraries for Open Knowledge: 23rd International
  Conference on Theory and Practice of Digital Libraries, TPDL 2019, Oslo,
  Norway, September 9-12, 2019, Proceedings 23}, pp. 296--311. Springer, 2019.

\bibitem{shah2002information}
U.~Shah, T.~Finin, A.~Joshi, R.~S. Cost, and J.~Matfield.
\newblock Information retrieval on the semantic web.
\newblock In {\em Proceedings of the eleventh international conference on
  Information and knowledge management}, pp. 461--468, 2002.

\bibitem{shen2023chatgpt}
Y.~Shen, L.~Heacock, J.~Elias, K.~D. Hentel, B.~Reig, G.~Shih, and L.~Moy.
\newblock Chatgpt and other large language models are double-edged swords,
  2023.

\bibitem{sheng2019cepv}
S.~Sheng, P.~Zhou, and X.~Wu.
\newblock Cepv: A tree structure information extraction and visualization tool
  for big knowledge graph.
\newblock In {\em 2019 IEEE International Conference on Big Knowledge (ICBK)},
  pp. 221--228. IEEE, 2019.

\bibitem{sievert2014ldavis}
C.~Sievert and K.~Shirley.
\newblock Ldavis: A method for visualizing and interpreting topics.
\newblock In {\em Proceedings of the workshop on interactive language learning,
  visualization, and interfaces}, pp. 63--70, 2014.

\bibitem{soylu2018optiquevqs}
A.~Soylu, E.~Kharlamov, D.~Zheleznyakov, E.~Jimenez-Ruiz, M.~Giese, M.~G.
  Skj{\ae}veland, D.~Hovland, R.~Schlatte, S.~Brandt, H.~Lie, et~al.
\newblock Optiquevqs: A visual query system over ontologies for industry.
\newblock {\em Semantic Web}, 9(5):627--660, 2018.

\bibitem{suchanek2007yago}
F.~M. Suchanek, G.~Kasneci, and G.~Weikum.
\newblock Yago: a core of semantic knowledge.
\newblock In {\em Proceedings of the 16th international conference on World
  Wide Web}, pp. 697--706, 2007.

\bibitem{8731539}
Y.~Sun, J.~Yu, and M.~Sarwat.
\newblock Demonstrating spindra: A geographic knowledge graph management
  system.
\newblock In {\em 2019 IEEE 35th International Conference on Data Engineering
  (ICDE)}, pp. 2044--2047, 2019. doi: {{%
10\hspace{.1pt}\discretionary{.}{%
}{.}\hspace{.4pt}1109\discretionary{/}{%
}{/}ICDE\hspace{.1pt}\discretionary{.}{%
}{.}\hspace{.4pt}2019\hspace{.1pt}\discretionary{.}{%
}{.}\hspace{.4pt}00235}}


\bibitem{tosi2021scikgraph}
M.~D.~L. Tosi and J.~C. Dos~Reis.
\newblock Scikgraph: A knowledge graph approach to structure a scientific
  field.
\newblock {\em Journal of Informetrics}, 15(1):101109, 2021.

\bibitem{tu2022phrasemap}
Y.~Tu, R.~Qiu, Y.-S. Wang, P.-Y. Yen, and H.-W. Shen.
\newblock Phrasemap: Attention-based keyphrases recommendation for information
  seeking.
\newblock {\em IEEE Transactions on Visualization and Computer Graphics}, 2022.

\bibitem{vargas2019rdf}
H.~Vargas, C.~Buil-Aranda, A.~Hogan, and C.~L{\'o}pez.
\newblock Rdf explorer: A visual sparql query builder.
\newblock In {\em The Semantic Web--ISWC 2019: 18th International Semantic Web
  Conference, Auckland, New Zealand, October 26--30, 2019, Proceedings, Part I
  18}, pp. 647--663. Springer, 2019.

\bibitem{waltl2018rule}
B.~Waltl, G.~Bonczek, and F.~Matthes.
\newblock Rule-based information extraction: Advantages, limitations, and
  perspectives.
\newblock {\em Jusletter IT (02 2018)}, 2018.

\bibitem{wang2020microsoft}
K.~Wang, Z.~Shen, C.~Huang, C.-H. Wu, Y.~Dong, and A.~Kanakia.
\newblock Microsoft academic graph: When experts are not enough.
\newblock {\em Quantitative Science Studies}, 1(1):396--413, 2020.

\bibitem{wang2019multi}
Z.~Wang, P.~Ng, X.~Ma, R.~Nallapati, and B.~Xiang.
\newblock Multi-passage bert: A globally normalized bert model for open-domain
  question answering.
\newblock {\em arXiv preprint arXiv:1908.08167}, 2019.

\bibitem{wei2020vision}
J.~Wei, S.~Han, and L.~Zou.
\newblock Vision-kg: topic-centric visualization system for summarizing
  knowledge graph.
\newblock In {\em Proceedings of the 13th International Conference on Web
  Search and Data Mining}, pp. 857--860, 2020.

\bibitem{wise2020covid}
C.~Wise, V.~N. Ioannidis, M.~R. Calvo, X.~Song, G.~Price, N.~Kulkarni,
  R.~Brand, P.~Bhatia, and G.~Karypis.
\newblock Covid-19 knowledge graph: accelerating information retrieval and
  discovery for scientific literature.
\newblock {\em arXiv preprint arXiv:2007.12731}, 2020.

\bibitem{yun2021knowledge}
W.~Yun, X.~Zhang, Z.~Li, H.~Liu, and M.~Han.
\newblock Knowledge modeling: A survey of processes and techniques.
\newblock {\em International Journal of Intelligent Systems}, 36(4):1686--1720,
  2021.

\end{thebibliography}

\end{document}


\maketitle
\section{SKG}
\begin{figure}
    \centering
    \includegraphics[width=\linewidth]{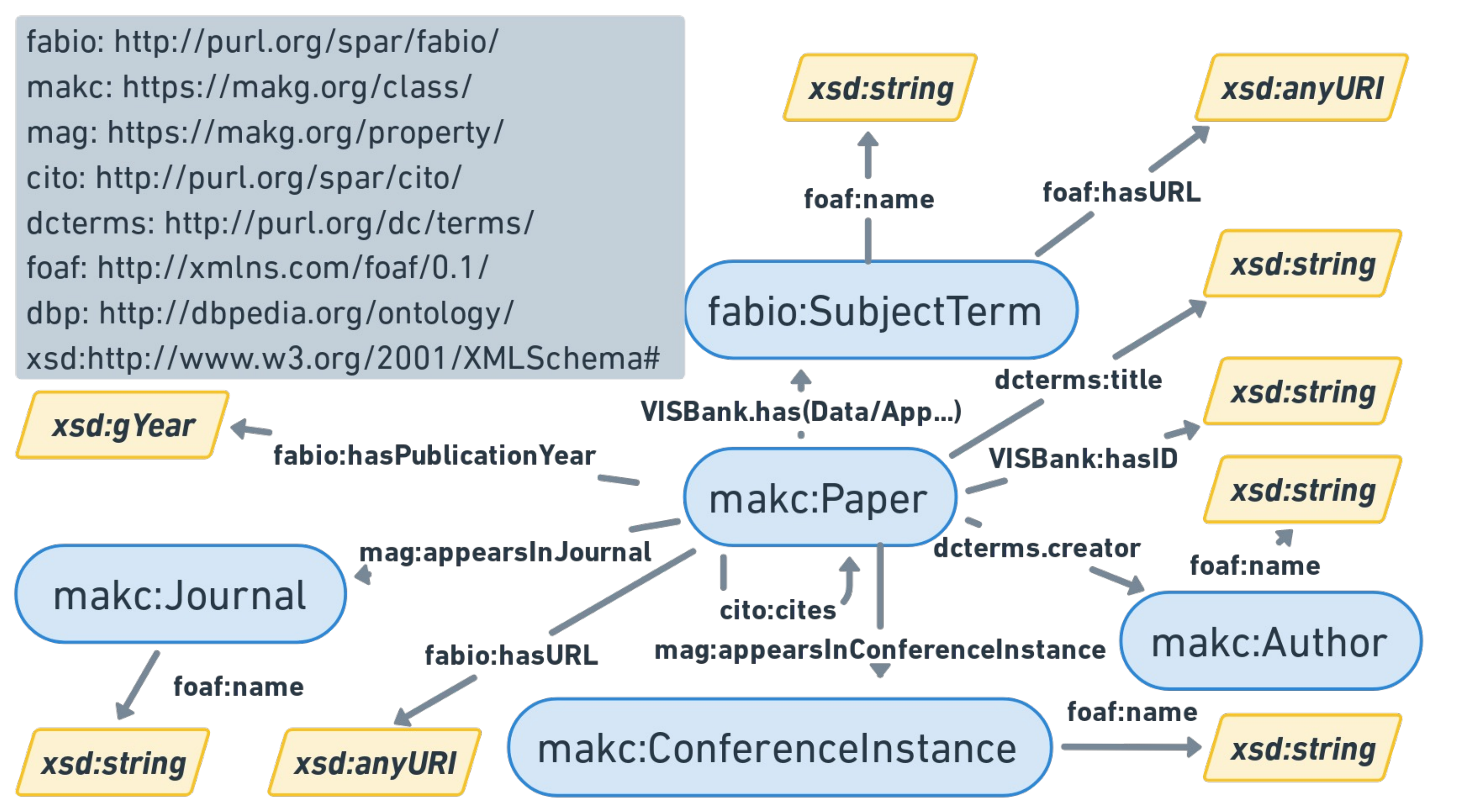}
    \caption{The ontology of the Semantic Knowledge Graph (SKG).}
    \label{fig:ontology}
\end{figure}
\section{Evaluation}
\subsection{Additional Exploration of Text Mining Field}
\begin{figure*}
    \centering
    \includegraphics[width=\textwidth]{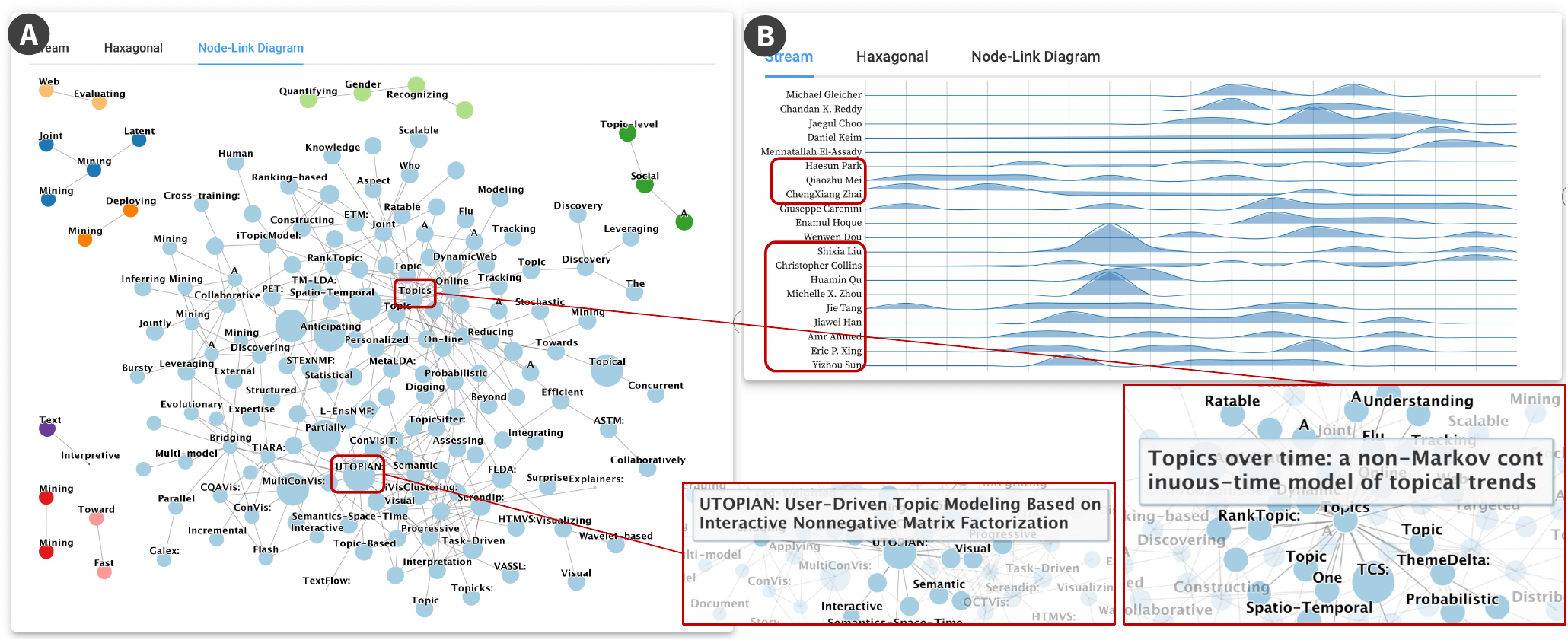}
    \caption{Caption}
    \label{fig:case_tm}
\end{figure*}
\subsubsection{Citation analysis}
 To achieve the citation relationships, she utilized a \emph{visualizer} (\autoref{fig:case_tm}D) and switched to the node-link diagram. 
 \autoref{fig:case_tm}D' illustrates the citation network, where she identified the biggest cluster focusing on topic modeling. This cluster contains the two most cited papers: one ~\cite{wang2006topics} proposes a \emph{Topics Over Time (TOT)} model that captures word co-occurrence pattern with time, while the other~\cite{choo2013utopian}, a highly cited paper in the visualization literature, analyzes how to incorporate user feedback into the topic modeling algorithms. 
 By hovering each node, related nodes pop up, assisting Emily in understanding which papers are in the intermediate section between visualization and technique to determine the most appropriate ones to read further.

\subsubsection{Temporal Analysis}
Emily also wanted to identify some top scholars in this area and keep track of their research work. To achieve this, she dragged a \emph{expander} (\autoref{fig:case_tm}G) to expand the graph by \emph{Author} entity and visualize the results in a stream graph (\autoref{fig:case_tm}H, H'). 
The stream chart displays the publications of the top 20 scholars from 2004 to 2020, revealing that each scholar has different active periods and patterns in this field. 
For instance, \emph{Shixia Liu} and \emph{Huaming Qu} had a surge in 2010, 
\emph{Jaegul Choo, Wenwen Dou} began their research in this area around 2012,
\emph{Jiawei Han}, \emph{Jie Tang}, \emph{Amr Ahmed} have demonstrated continuous research interests in this field.

\section{VISBank}
\subsection{Venues}
\begin{itemize}
    \item 
\end{itemize}

\subsection{VISBank}
\section{NER Models}
\subsection{Training details}
\begin{table*}[]
\centering
\begin{tabular}{lllllll}
\hline
 & Huggingface Name & Loss & Precision & Recall & F1 & Accuracy \\ \hline
BERT-base & Yamei/bert-base-uncased-v10-ES-ner & 0.3872 & 0.6667 & 0.7066 & 0.6861 & 0.9163 \\
BERT-large & Yamei/bert-large-uncased-v10-ES-ner & 0.5047 & 0.6503 & 0.7107 & 0.6792 & 0.9157 \\
DistilBERT & Yamei/distilbert-base-uncased-v10-ES-ner & \textbf{0.3587} & 0.6550 & 0.6983 & 0.676 & 0.9157 \\
SciBERT & Yamei/scibert\_scivocab\_uncased-v10-ES-ner & 0.4185 & 0.6897 & \textbf{0.7616} & \textbf{0.7239} & \textbf{0.9263} \\
RoBERTa & Yamei/xlm-roberta-large-v10-ES-ner & 0.4403 & \textbf{0.6980} & 0.7355 & 0.7163 & 0.9175 \\ \hline
\end{tabular}
\caption{TBA}
\label{tab:model}
\end{table*}
We fine-tuned five pre-trained language models for Named Entity Recognition. 
Our fine-tuning process utilized a learning rate of 2e-5, which is a common choice for large pre-trained language models as it balances the need for rapid convergence with the risk of overfitting. We also implemented a weight decay of 0.01 as a penalty term for the loss function to prevent overfitting. In addition, we employed an early-stop mechanism that monitored the models' f1 score on a validation set during training and stopped when the performance no longer improved. We saved the best model at the end of the training, and its performance, along with the other models, is presented in \autoref{tab:model}.

\subsection{Model Inference}
\begin{figure}
    \centering
    \includegraphics[width=\linewidth]{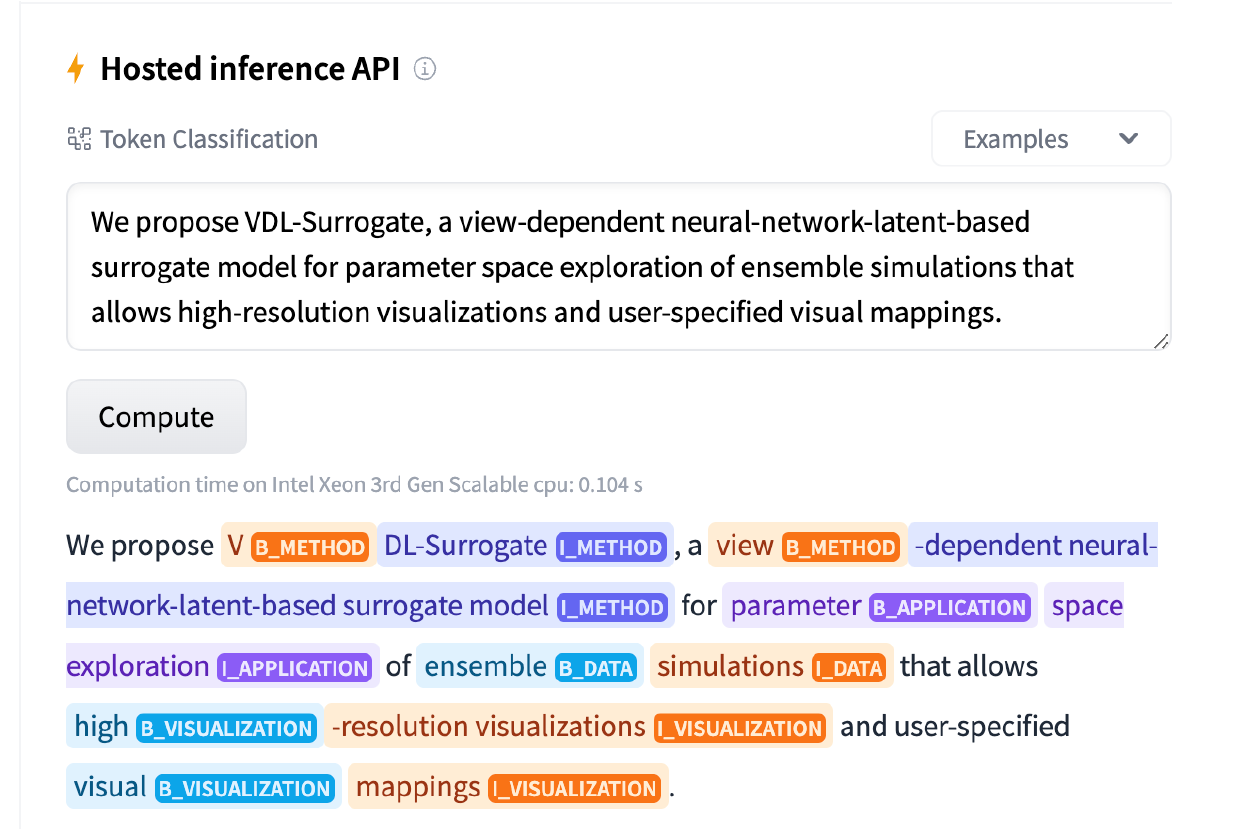}
    \caption{One example of testing our NER model using hosted inference API by Huggingface.}
    \label{fig:huggingface}
\end{figure}

We developed an NER model specifically designed to identify entities in the visualization literature. The models are publicly available on the Huggingface hub and can be easily tested using their inherent hosted inference API, as illustrated in Figure 1. 
The B and O indicate the beginning token and the following tokens of each entity. For instance, in the \autoref{fig:huggingface}, ``VDL-Surrogate'' is a method entity, and ``view-dependent neural-network-latent-based surrogate model'' is another method entity that the model can identify.
Moreover, we provide an example of how to import our trained NER model in Python:
\begin{lstlisting}[language=Python, caption=Python example]
from transformers import pipeline
text = """
"""
ner_model = pipeline("ner", model="Yamei/xlm-roberta-large_NER_VISBank")
entities = ner_model(text)
\end{lstlisting}